\documentclass[%
 reprint,
 amsmath,amssymb,
 aps,
 prl,
]{revtex4-1}

\usepackage{graphicx}
\usepackage{dcolumn}
\usepackage{bm}
\usepackage{float}
\usepackage{array}

\usepackage{braket}
\usepackage{amsmath}
\usepackage{amssymb}

\usepackage{eucal}

\usepackage{epstopdf}

\begin{document}

\preprint{APS/123-QED}

\title{Complementary First and Second Derivative Methods for Ansatz Optimization in Variational Monte Carlo}

\author{Leon Otis$^1$}

\author{Eric Neuscamman$^{2,3,}$}%
\email{eneuscamman@berkeley.edu}

\affiliation{
${}^1$Department of Physics, University of California, Berkeley, CA, 94720, USA \\
${}^2$Department of Chemistry, University of California, Berkeley, CA, 94720, USA \\
${}^3$Chemical Sciences Division, Lawrence Berkeley National Laboratory, Berkeley, CA, 94720, USA
}

\date{\today}

\begin{abstract}
We present a comparison between a number of
recently introduced low-memory wave function optimization
methods for variational Monte Carlo in which we find that
first and second derivative methods possess strongly complementary
relative advantages.
While we find that low-memory variants of the linear method
are vastly more efficient at bringing wave functions with
disparate types of nonlinear parameters to the vicinity of the
energy minimum, accelerated descent
approaches are then able to locate the precise minimum with
less bias and lower statistical uncertainty.
By constructing a simple hybrid approach that combines these
methodologies, we show that all of these advantages can be
had at once when simultaneously optimizing large determinant
expansions, molecular orbital shapes, traditional Jastrow
correlation factors, and more nonlinear
many-electron Jastrow factors.
\end{abstract}

\maketitle

\section{Introduction}

The practical utility of widely used methods in electronic structure
theory is in large part determined by the optimization algorithms
they rely on.
This basic theme has been repeated throughout the history of
quantum chemistry, with methods as fundamental as Hartree-Fock
theory becoming dramatically more useful with the development
of superior solution methods such as the direct inversion of
the iterative subspace. \cite{pulay1982diis}
Similar transformations have been seen in configuration interaction
(CI) theory thanks to Davidson's method, \cite{davidson1975}
in the density matrix renormalization group (DMRG) approach
thanks to (among other innovations) the noise algorithm, \cite{white2005noise}
and in many other methods besides.
As in the case of DMRG, it is usually not so simple as a single innovation
in the numerical methods that transforms a theory from a promising
proof of concept into a robust computational tool.
Instead, such tools often arise as the result of a series of innovations,
that, once combined, fit together in a way that makes them more than
the sum of their parts.

In the context of quantum Monte Carlo (QMC), and more specifically
in its variational (VMC) formulation, the introduction of the
linear method (LM) for trial function optimization marked a large
step forward along the path to practical utility and reliability.
\cite{Umrigar2007}
However, recent research has revealed multiple options for
bypassing the LM's memory bottleneck, making clear that there is
still a great deal of distance to cover in the maturation of
VMC numerical methods.
Some of these approaches
\cite{Neuscamman2012,Zhao2017}
depend, like the LM itself, on knowing
at least some information about energy second derivatives, but by
avoiding the construction of full Hessian-sized matrices they
achieve dramatically lower memory footprints compared to the LM.
Other even more recent approaches, most of which can be classified as
accelerated descent (AD) methods,
\cite{Schwarz2017,Sabzevari2018,Mahajan2019}
avoid second derivative information entirely and are thus even
more memory efficient, relying instead on a
limited knowledge of the optimization's history of energy
first derivatives or in one case just the signs of
these derivatives. \cite{Luo2018}
In the present study, we explore the relative advantages of these
first and second derivative approaches and find that,
when combined, they offer a highly complementary optimization
strategy that appears both more robust and more efficient
than either class of methods is on its own.

The ability to optimize larger and more complicated wave function
forms is becoming increasingly relevant due to rapid
progress in other areas of VMC methodology.
The introduction of the table method \cite{Clark2011,Morales2012}
has increased the size of CI expansions that can be
handled by more than an order of magnitude, and expansion
lengths beyond 10,000 determinants are no longer unusual.
A recent improvement to the table method \cite{Filippi2016,Assaraf2017}
now allows the molecular orbital basis to be optimized
efficiently in the presence of these large expansions,
while the resurgence of interest in selected CI methods
\cite{evangelista2016adaptive,Holmes2016,tubman2016asci,
      Sharma2017,loos2018scipt,zimmerman2018sshci}
has provided a convenient route to their construction.
In addition to these CI-based advances, other wave function
innovations have also led to growing demands on
VMC optimization methods.
Increasingly sophisticated correlation factors,
such as those used in Hilbert space approaches
\cite{changlani2009cps,mezzacapo2009eps,neuscamman2012magnet,
      Neuscamman2012,Schwarz2017,Sabzevari2018,Mahajan2019}
as well as a steady stream of developments in real space
\cite{Casula2004,Sorella2007,LopezRios2012,Luchow2015,Goetz2017,Goetz2018}
have also raised the demand for optimization approaches
that can deal with large numbers of highly nonlinear parameters.
Although less thoroughly explored, the treatment of correlation effects
via back flow transformations also continues to receive attention
and create new optimization challenges. \cite{Holzmann2015iterbackflow,Luo2018}
Finally, in addition to these increases in ansatz sophistication,
recent interest in using excited state variational principles
to expand QMC's excited state capabilities has led to its own
collection of optimization difficulties.
\cite{Zhao2016,neuscamman2016varqmc,blunt2017ct,Shea2017,
      robinson2017vm,blunt2018excited,Flores2019}

By supporting these various advances in QMC methodology, improved
VMC optimization methods have the potential for large impacts
in diverse areas of chemistry and solid state physics.
Work on lattice models, for example, continues to push the
boundaries on how approximate wave functions are defined.
\cite{troyer2017rbm,Kochkov2018}
In the area of molecular excited states, QMC methods offer
promising new routes to high-accuracy treatments of both
double excitations \cite{Zhao2016,Neuscamman2016}
and charge transfer excitations, \cite{blunt2017ct,Flores2019}
both of which continue to challenge conventional
quantum chemistry methods.
In QMC's traditional area of simulating real solids, applications
of both VMC and projector Monte Carlo would benefit immediately
from the ability to prepare more sophisticated trial
wave functions.
\cite{Foulkes2001,morales2018qfqmcnio,zhao2018gaps}
Diffusion Monte Carlo (DMC) in particular would achieve higher accuracy using 
the better nodal surfaces determined by well-optimized ansatzes from VMC.
More generally, the ability of QMC to combine treatments
of weak and strong electron correlation effects within a robust
variational framework that operates near the basis set limit
makes it a powerful general-purpose
approach for difficult molecular and materials problems where
high accuracy is necessary.
By increasing the size and complexity of systems that fall into
its purview, improvements in QMC wave function optimization methods
therefore have the potential to move electronic structure simulation
forward on a number of fronts.

The present study seeks to aid in this endeavor by focusing on
the relative advantages of recently developed low-memory
first and second derivative methods in VMC and in
particular on how they can be used to complement each other.
Unlike deterministic optimizations, in which second derivative
methods are typically preferred so long as they are affordable,
the situation is less straightforward when the objective function
and its derivatives are statistically uncertain.
One major concern is that, in practice, it can be more difficult
to achieve low-uncertainty estimates of the second derivative
terms that appear in the LM and its descendants.
While this issue can be mitigated by the use of alternative
approaches to importance sampling,
these can increase uncertainty in the energy due to the loss
of the zero-variance principle.
Thus, as we will demonstrate, statistical precision tends to
be higher when using AD methods, which is an advantage on top
of their ability to converge to the minimum without the bias
that arises from the LM's highly nonlinear matrix diagonalization.
However, we will also see that in order to enjoy the advantages
of a tighter and less biased final convergence,
AD methods must first reach the vicinity of the minimum.
For this task, we find that the LM and its low-memory variants
outperform all of the first derivative methods that we tested,
especially for optimizations in which the wave function contains
different classes of parameters that vary greatly in their
nonlinear character and how they couple to each other.
Happily, we will see that a hybrid approach --- in which
AD and low-memory LM optimization steps are interwoven ---
excels both at reaching the vicinity of the minimum and
producing unbiased final energies while simultaneously
maintaining a high degree of statistical efficiency.

\section{Theory}

\subsection{Variational Monte Carlo}

VMC combines the variational principle of quantum mechanics with Monte Carlo evaluation of high dimensional integrals.\cite{Umrigar2015}
To study the ground state of a system, we pick a trial wave function $\Psi$ of some particular form and seek to minimize its energy expectation value.
\begin{equation}
    E(\Psi) = \frac{\Braket{\Psi | H | \Psi}}{\Braket{\Psi | \Psi}}
\end{equation}

In the language of mathematical optimization, $E(\Psi)$ is an example of an objective function or cost function.
For a typical system with $N$ electrons, this expression contains integrals over $3N$ position space coordinates which for some wave functions can only be evaluated efficiently through Monte Carlo sampling rather than quadrature methods.
We rewrite the energy as
\begin{align}
\notag
   E &= \frac{\int d \mathbf{R}\Psi (\mathbf{R}) H \Psi(\mathbf{R})}{\int d \mathbf{R}\Psi (\mathbf{R})^2}
      = \frac{\int d \mathbf{R}\Psi (\mathbf{R})^2 E_L (\mathbf{R})}{\int d \mathbf{R}\Psi (\mathbf{R})^2} \\
     &= \int d \mathbf{R} \rho (\mathbf{R})E_L (\mathbf{R})
\label{eqn:energy}
\end{align}
where $E_L (\mathbf{R}) = \frac{H \Psi (\mathbf{R})}{\Psi (\mathbf{R})}$ is the local energy and $\rho (\mathbf{R}) = \frac{\Psi(\mathbf{R})^2}{\int d \mathbf{R} \Psi (\mathbf{R})^2}$ is the probability density.
The zero-variance principle\cite{Assaraf1999} makes $\rho (\mathbf{R})$ the most common choice of probability distribution for obtaining samples, but it is not the only option.
For effective estimation of quantities beside the energy, such as the LM matrix elements, other
importance sampling functions are often preferred.
\cite{trail2008a,trail2008b,robinson2017vm}
In our LM and blocked LM calculations in this study,
we employ the importance sampling function
(and the appropriately modified statistical estimate formulas
\cite{Flores2019})
\begin{align}
\label{eqn:is}
    |\Phi|^2 \equiv
    |\Psi|^2 + \frac{\epsilon}{M}
    \hspace{0.5mm} \sum_{I} |D_I|^2
\end{align}
in which the $D_I$ are the $M$ different $S_z$-conserving single
excitations relative to the closed shell reference determinant.
The logic behind this choice is that it puts some weight on
configurations that are highly relevant for the orbital rotation
parameters' wave function derivatives, as small orbital
rotations can be approximated via the addition of singles.
We find that this importance sampling function substantially
reduces the uncertainty of the LM matrix elements corresponding
to orbital rotations, which in turn helps reduce the update
step uncertainty.
For AD, we simply use traditional $|\Psi|^2$ importance sampling
as in equation \ref{eqn:energy}.

By the variational principle, we are guaranteed that $E$ is an upper bound on the true ground state energy. 
Given some set of adjustable parameters in the functional form of $\Psi$, we expect that values of those parameters that yield a lower value of $E$ to correspond to a wave function that is closer to the ground state. 
One could then imagine the abstract space produced by the possible values of all variational parameters.
The set of optimal parameter values that specify the wave function expression which minimizes $E$ can be taken as a point in this space labeled by the vector $\mathbf{p^*}$.
In general, the initial choice for parameters will not be at this energy minimum point, but at some other point $\mathbf{p_0}$.
The problem of determining the best wave function in VMC calculations then relies on an optimization algorithm for finding $\mathbf{p^*}$ after starting from $\mathbf{p_0}$.

Within this framework, one of the most important considerations is that the optimization is inherently stochastic due to the introduction of noise through the Monte Carlo evaluation of the integral in equation \ref{eqn:energy}.
This forms a contrast with many other methods in electronic structure theory including Hartree-Fock, CI, and coupled cluster where various deterministic optimization schemes predominate.\cite{Helgaker2000}
Many of the algorithms commonly encountered in a deterministic quantum chemistry context such as steepest descent and the Newton method, have been adapted for use in VMC.\cite{Harju1997,Lin2000,Lee2005,Sorella2005,Umrigar2005}
However, there is now a need to be robust to the presence of noise.
Historically, errors due to finite sampling led to numerical instabilities that prompted interest in minimizing variance\cite{Kent1999,Umrigar1988} instead of energy, but later optimization developments have sought to mitigate this issue and in this paper we only consider energy minimization.
As we will now discuss in their respective sections, both the LM and gradient descent approaches possess features that enable them to operate stably in a stochastic setting.

\subsection{The Linear Method}
The LM\cite{Nightingale2001,Umrigar2007} begins with a first order Taylor expansion of the wave function.
For a set of variational parameters given by vector $\mathbf{p}$, we have
\begin{equation}
\label{eqn:lmTaylor}
    \Psi (\mathbf{p}) = \Psi_0 + \sum_i \Delta p_i \Psi_i
\end{equation}
where $\Psi_i = \frac{\partial \Psi(\mathbf{p})}{\partial p_i}$ and $\Psi_0$ is the wave function at the current parameter values.

Finding the optimal changes to the parameters amounts to solving the generalized eigenvalue problem 
\begin{equation}
\label{eqn:lmEigen}
    H\Vec{c} = E S \Vec{c}
\end{equation}
 in the basis of the wave function and its first order parameter derivatives $\{\Psi_0,\Psi_1,\Psi_2,...\}$.
$H$ and $S$ are the Hamiltonian and overlap matrices in this basis with elements 
\begin{equation}
\label{eqn:lmH}
    H_{ij} = \Braket{\Psi_i | H | \Psi_j} 
\end{equation}
\begin{equation}
\label{eqn:lmS}
    S_{ij} = \Braket{\Psi_i  | \Psi_j}
\end{equation}
The matrix diagonalization to solve this eigenproblem for eigenvector $\Vec{c} = (1,\Delta \mathbf{p})$ then yields the updated parameter values $\mathbf{p_1} = \mathbf{p_0} + \Delta \mathbf{p}$.
As the matrices $\bm{H}$ and $\bm{S}$ both contain a subset
of the second derivative terms that would be present in a 
Newton-Raphson approach, \cite{Toulouse2007}
the LM is most naturally categorized as a second-derivative
method, and it certainly shares Newton-Raphson's difficulties
with regards to dealing with matrices whose dimension
grows as the number of variables.

For practical use with finite sampling, the LM must be stabilized to prevent unwisely large steps in parameter space.
This is accomplished by adding shift values\cite{Umrigar2007} to the matrix diagonal that effectively act as a trust radius scheme similar to those used with the Newton method.
In our implementation, the Hamiltonian is modified with two shift values meant to address distinct potential problems in the optimization.\cite{Kim2018}
\begin{equation}
\label{eqn:lmShift}
    \mathbf{H} \xrightarrow[]{} \mathbf{H} + c_I \mathbf{A} + c_S \mathbf{B}
\end{equation}

The matrix elements of $\mathbf{A}$ are given by $A_{ij} = \delta_{ij}(1-\delta_{i0})$ so that the shift $c_I$ effectively gives an energy penalty to directions of change from the current wave function.\cite{Umrigar2007}
The second shift is intended to address problems that may arise if some wave function derivatives have norms that differ by orders of magnitude. 
In this situation, the single shift value $c_I$ is insufficient to preserve a quick yet stable optimization.
For a parameter with a large derivative norm, a sufficiently high value of $c_I$ might prevent an excessively large change in its value.
However, all other parameter directions with smaller derivative norms will be so heavily penalized by the large value of $c_I$ that those parameters become effectively fixed.
The purpose of the second $c_S \mathbf{B}$ term is to retain important flexibility in other parameter directions. 
We can write the matrix $\mathbf{B}$ as
\begin{equation}
\label{eqn:bMat}
\mathbf{B} = (\mathbf{Q^T})^{-1} \mathbf{T} \mathbf{Q}^{-1}
\end{equation}
where
\begin{equation}
\label{eqn:qMat}
    Q_{ij} = \delta_{ij} -\delta_{i0}(1-\delta_{j0})S_{0j}
\end{equation}
and 
\begin{equation}
\label{eqn:tMat}
    T_{ij} = (1-\delta_{i0}\delta_{j0})[\mathbf{Q^T}\mathbf{S}\mathbf{Q}]_{ij}
\end{equation}
The matrix $\mathbf{Q}$ provides a transformation to a basis where all update directions are orthogonal to the current wave function and the matrix $\mathbf{T}$ is the overlap matrix in this basis. 
The optimal choice of shift parameters $c_I$ and $c_S$ may depend on the particular optimization problem.
In our implementation, an adaptive scheme adjusts the shifts on each iteration by comparing the energies calculated through correlated sampling on three different sets of shift values and choosing whichever shifts produced the lowest energy.

The LM has been successfully applied to a
variety of systems to prepare good trial wave functions
for DMC.
\cite{Umrigar2007,Toulouse2007,Toulouse2008,Brown2007,Petruzielo2012,Goetz2017,Goetz2018}
It has also been used in the variational optimization of a recent functional for targeting excited states.\cite{Zhao2016,Shea2017,Flores2019}
However, it possesses a number of limitations, most notably a memory cost that scales with the square of the number of optimizable parameters due to the matrices it builds. 
This memory cost currently confines routine use of the LM to less than roughly 10,000 parameters though exceptional calculations with up to about 16,000 have been made.\cite{Clark2011}
Another shortcoming is the nonlinear bias of the LM.
We are evaluating the elements of the Hamiltonian and overlap matrices stochastically and have a nonlinear relationship between them and our energy through the generally high order characteristic polynomial of the eigenvalue problem of equation \ref{eqn:lmEigen}.
As a result, we in general expect the LM to converge to a point in parameter space slightly offset from the true minimum.
This nonlinear bias has been studied for the LM in Hilbert space\cite{Zhao2016a} and a similar issue arises in the context of Full Configuration Interaction QMC.\cite{Blunt2018}
Both the memory constraint and the nonlinear bias of the LM become more severe for ansatzes with larger numbers of variational parameters, which spurs the search for potential alternatives.
One approach suggested for memory reduction is to employ Krylov subspace methods for Eq. 4 to avoid building matrices, but it requires a drastically higher sampling effort due to the need for many matrix-vector multiplications and so we do not pursue the approach here.\cite{Neuscamman2012}

\subsection{Blocked Linear Method}
One recent approach to bypassing the memory bottleneck is known as the blocked linear method (BLM).\cite{Zhao2017}
The first step of the algorithm is to divide the full set of parameters into $N_b$ blocks. 
Next, a LM-style matrix diagonalization is carried out within each block and some number $N_k$ of the resulting eigenvectors from the blocks are retained as good directions for constructing an approximation for the overall best update direction in the full parameter space.
For a particular block of variables, the wave function expansion in the LM is given by
\begin{equation}
\label{eqn:blmBlock}
    \ket{\Psi_b} = \ket{\Psi_0} + \sum_{i=1}^{M_b} c_i \ket{\Psi^i} 
\end{equation}
where $\ket{\Psi^i}$ is the wave function derivative with respect to the $i$th variable in the block, $M_b$ is the number of variables in the block, and $\ket{\Psi_0}$ the current wave function as in the normal LM.
We can perform the same matrix diagonalization done in the LM, only with parameters outside the block fixed.
This yields a set of eigenvectors that we can use to construct another approximate expansion of the original wave function.
We can construct a matrix $\mathbf{B}$ using the $N_k$ eigenvectors with the lowest eigenvalues from each block and write a new expansion
\begin{equation}
\label{eqn:blmNewBlock}
    \ket{\Tilde{\Psi}} = \alpha \ket{\Psi_0} + \sum_{k=1}^{N_b} \sum_{j=1}^{N_k} A_{kj} \sum_{i=1}^{M_b} B_{ji}^{(b)} \ket{\Psi^{i,b}}
\end{equation}
Having now pre-identified important directions within each block,
the idea is that a subsequent LM-style diagonalization
in the basis of these good directions (which yields
the coefficients $A_{kj}$) should still provide a good
update direction when re-expressed in the full parameter space.

In order to help retain most of the accuracy of the traditional LM, the first stage of the BLM computation includes $N_o$ other good directions that are used to supply the current block's diagonalization with information about how its variables are likely to couple to those in other blocks.
In practice, important out-of-block directions are obtained by
keeping a history of previous iterations' updates
as the optimization progresses.
We can rewrite the one block expansion introduced in equation \ref{eqn:blmBlock} as
\begin{equation}
\label{eqn:blmHistory}
    \ket{\Psi_b} = \ket{\Psi_0} + \sum_{i=1}^{M_b} c_i \ket{\Psi^i} + \sum_{j=1}^{N_o} \sum_{k=1,k\neq b}^{N_b} d_{jk} \ket{\Theta_{jk}}
\end{equation}
where we take the
\begin{equation}
\label{eqn:blmTheta}
    \ket{\Theta_{jk}} = \sum_{l=1}^{M_k} C_{jkl} \ket{\Psi^{l,k}}
\end{equation}
as the linear combinations of wave function derivatives from other blocks that were identified as important based on previous iterations' updates.
The additional term in the expansion allows us to account for couplings between variables in different blocks and enable the construction of a better space for the second diagonalization.
We assemble the matrix $\mathbf{B}$ and $\ket{\Tilde{\Psi}}$ and then seek to minimize $\frac{\Braket{\Tilde{\Psi} | H | \Tilde{\Psi}}}{\Braket{\Tilde{\Psi}|\Tilde{\Psi}}}$ with respect to variational parameters $\alpha$ and $A_{kj}$ in our BLM wave function expansion in equation \ref{eqn:blmNewBlock}.

\begin{figure}
    \centering
    \includegraphics[width=8.3cm]{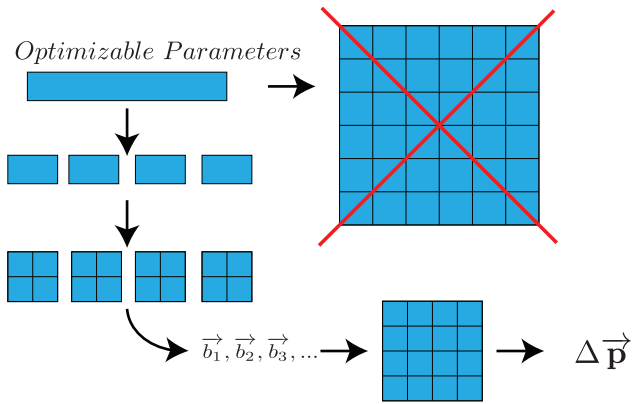}
    \caption{Flowchart depicting steps in the BLM algorithm to arrive at a parameter update.}
    \label{fig:blmchart}
\end{figure}

Figure \ref{fig:blmchart} portrays the algorithmic steps described above. Some number of parameters too large to be handled by the standard LM is divided among different blocks whose diagonalizations produce the vectors $\Vec{b_i}$ for the construction of the space of the second diagonalization that produces the parameter update.
The BLM can be thought of as achieving memory savings in the use of smaller matrices at the cost of having to run over the sample twice when the traditional LM must run over it just once.
A more extensive description of the BLM and its precise memory usage can be found in its original paper.\cite{Zhao2017}

We divide parameters evenly among blocks, but one could implement the use of tailored blocks of varying sizes. 
It is advisable to choose the block size to be large enough to keep important parameters of the same type, such as all of those for a Jastrow factor, within the same block.
This enables the expected strong coupling between them to be handled more accurately by the LM-style diagonalization within that block.
While the BLM has been successfully applied up to about 25,000 parameters and found to closely reproduce the results of the standard LM, \cite{Zhao2017}
it remains a relatively new method, and the present study will provide additional data on its efficacy.

\subsection{Gradient Descent Methods}
In the last few years, increasing attention\cite{Schwarz2017,Schwarz2017a,Sabzevari2018,Luo2018,Mahajan2019} has been paid to optimization methods that use only first derivatives to minimize a Lagrangian of the form 
\begin{equation}
\label{eqn:lagrangian}
    \mathcal{L}( \Psi(\mathbf{p})) = \Braket{\Psi | H | \Psi} - \mu(\Braket{\Psi | \Psi} - 1) 
\end{equation}
where $\mu$ is a Lagrange multiplier and, in practice, a moving average of the local energy.
There is no need to solve an eigenvalue problem as in the LM and the memory cost of these approaches scales linearly with the number of parameters.
We also note that the stochastic evaluation of derivatives of this Lagrangian will lead to a smaller nonlinear bias compared to what is encountered in the LM.
While there is some nonlinearity present in the product $\mu \Braket{\Psi | \Psi}$, it is mild compared to the high order polynomials encountered in the solution of the LM eigenvalue problem and can be avoided entirely if desired through modest amounts of extra sampling.
Minimization of this Lagrangian targets the ground state, but excited states can similarly be targeted with these optimization algorithms merely by using derivatives of one of the excited state functionals that have been developed.\cite{Choi1970,Ye2017,Zhao2016}

The simplest method in this category is the steepest descent algorithm.
\begin{equation}
\label{eqn:steepest}
    p_i^{k+1} = p_i^k - \eta_k \frac{\partial \mathcal{L}(\mathbf{p})}{\partial p_i}
\end{equation}
In this case, the value of each parameter on the $k+1$'th step is found simply by subtracting the statistically uncertain parameter derivative times a step size $\eta_k$.
The step size can be taken as constant over all steps in the simplest case, but rigorous proofs on the convergence of stochastic gradient descent (SGD) rely on decaying step sizes satisfying $\sum_{k} \eta_k = \infty$ and $\sum_{k} \eta_k^2 < \infty$.\cite{Bottou2012}

It may be worth briefly commenting that the typical formulation of stochastic gradient descent as seen in the machine learning and mathematical optimization literature is slightly different from what we use here within VMC.
In a common machine learning scenario,\cite{Bottou2012} one has a training set of input data $\{x_1,x_2,...,x_n\}$ and corresponding outputs $\{y_1,y_2,...,y_n\}$ and wishes to minimize a loss function $Q(x,y;w)$ that measures the error produced by a model $f_w(x)$, which predicts $\widetilde{y}_i$ given $x_i$ and is parameterized by variables $w$.
For this setting, the SGD algorithm refers to evaluating the gradient of $Q$ with a randomly chosen pair $(x_j,y_j)$ from the given data set and then computing the parameter update according to $w_{k+1} = w_k - \eta_k \nabla_w Q(x_j,y_j)$.
For our VMC optimization, we are dealing with a noisy gradient similar to what occurs in this machine learning problem, but the source of our noise is somewhat different and lies in our means of evaluating the underlying 3N dimensional integrals within our Lagrangian derivatives.
Another important distinction is that in machine learning applications, complete convergence to the minimum is in fact undesirable because it will overfit the model to the training data and degrade its performance on new sets of test inputs.
Much as SGD provides a computational speed up for machine learning problems, we are also able to operate gradient descent methods at a cheap per-iteration cost because we need only a modest number of samples to evaluate sufficiently precise Lagrangian derivatives compared to the Hamiltonian and overlap matrices in the LM.
However, unlike the machine learning case, we do want to come as close as possible to the true minimum, and we will see that even reaching the vicinity of the minimum can be difficult for descent methods when typical VMC initial guesses are employed.

\begin{figure}
    \centering
    \includegraphics[width=8.3cm]{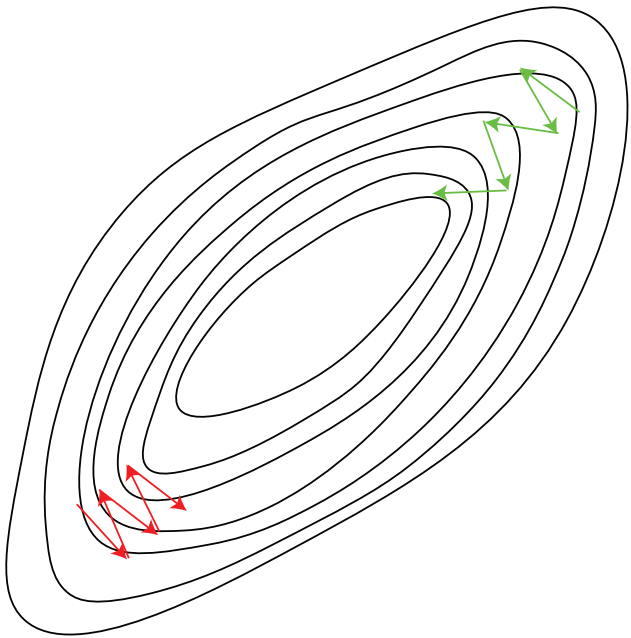}
    \caption{Illustration of the difficulty faced by steepest descent in red on the lower left with its slow approach to the minimum. Accelerated descent in green on the upper right is able to progress more rapidly to the minimum with its memory of previous gradients.}
    \label{fig:narrowvalley}
\end{figure}

While steepest descent can be guaranteed to eventually reach the minimum of the Lagrangian even in a stochastic setting, its asymptotic convergence is very slow.
For some intuition, one could imagine the landscape of the Lagrangian's values forming a very narrow valley near the true minimum.
In this situation, steepest descent would produce parameter updates mostly back and forth along the sides of the valley with little improvement of parameter values in the direction directly toward the minimum.
Due to the limitations of steepest descent, a number of other flavors of accelerated gradient descent (AD) have been developed that include a momentum term with information on previous values of the gradient.
As illustrated in Figure \ref{fig:narrowvalley}, the general intuition is that this additional term provides some memory of the progression along narrow valleys that steepest descent lacks and thereby achieves swifter convergence.
In addition, there are multiple schemes for adaptively varying the step sizes used in a manner that draws on the particular derivative values for each individual parameter as the optimization progresses. 
These methods have recently been applied successfully to Hilbert space QMC. In this study, we work in real space and investigate a combination of Nesterov momentum with RMSprop as presented by the Booth group \cite{Schwarz2017,Schwarz2017a}, a method using random step sizes from the Clark group\cite{Luo2018}, AMSGrad, recently used by the Sharma group\cite{Sabzevari2018,Mahajan2019}, as well as the ADAM optimizer\cite{Kingma}.

We now lay out the precise expressions for each of these methods in turn. The RMSprop algorithm used by Booth and co-workers is given by the following recurrence relations.\cite{Schwarz2017}
\begin{equation}
\label{eqn:rmspropMomentum}
    p_i^{k+1} = (1-\gamma_k)q_i^{k+1} - \gamma_k q_i^k
\end{equation}
\begin{equation}
\label{eqn:rmspropUpdate}
   q_i^{k+1} = p_i^k - \tau_k \frac{\partial \mathcal{L}(\mathbf{p})}{\partial p_i}
\end{equation}
\begin{equation}
\label{eqn:rmspropRecur}
\lambda_0=0 \hspace{7mm}
\lambda_k = \frac{1}{2} + \frac{1}{2}\sqrt{1+4\lambda_{k-1}^2} \hspace{7mm}
\gamma_k = \frac{1-\lambda_k}{\lambda_{k+1}}
\end{equation}
\begin{equation}
\label{eqn:rmspropStep}
    \tau_k = \frac{\eta}{\sqrt{E[(\frac{\partial\mathcal{L}}{\partial p_i})^2]^{(k)}} + \epsilon}
\end{equation}
\begin{equation}
\label{eqn:rmspropAvg}
    E[(\partial \mathcal{L})^2]^{(k)} = \rho E\left[\left(\frac{\partial\mathcal{L}}{\partial p_i}\right)^2\right]^{(k-1)} + (1-\rho)\left(\frac{\partial\mathcal{L}}{\partial p_i}\right)^2
\end{equation}
Above, $p_i^k$ denotes the value of the $i$th parameter on the $k$th step of the optimization, $\tau_k$ is a step size that is adaptively adjusted according to the RMSprop algorithm in equations \ref{eqn:rmspropStep} and \ref{eqn:rmspropAvg}.
The running average of the square of parameter derivatives in the denominator of $\tau_k$ allows for the step size to decrease when the derivative is large, which should hedge against the possibility of taking excessively large steps.
Conversely, a smaller denominator when the derivative is small allows for larger steps to be taken.
The weighting in the running average is controlled by a factor $\rho$ that can be thought of as the amount of memory retained of past gradients for adjusting $\tau_k$, and $\eta$ again denotes the chosen initial step size.
In order to avoid possible singularities when the gradient is very close to zero, a small positive number $\epsilon$ is included in the denominator of $\tau_k$.
Equation \ref{eqn:rmspropMomentum} shows the momentum effect in which the update for the parameter on the $k+1$ step depends on the update from the previous step as well as the current gradient.
We also follow the Booth group in applying a damping factor to the momentum by replacing $\gamma_k$ with $\gamma_k e^{-(\frac{1}{d})(k-1)}$.
The quantity $d$ effectively controls how quickly the momentum is turned off, which eventually turns the algorithm into SGD.
The values of $d$,$\eta$, $\rho$, and $\epsilon$ may all be chosen by the user of the algorithm and are known as hyperparameters in the machine learning literature.
In the results we present using this method, we have used $d=100$, $\rho = .9$ and $\epsilon = 10^{-8}$.
We have found adjusting these hyperparameters has relatively little influence on optimization performance compared to choices for step size $\eta$, but their influence could be explored more systematically.

The Clark group's algorithm takes a far simpler form
\begin{equation}
\label{eqn:randomStep}
    p_i^{k+1} = p_i^k - \alpha \eta \frac{\left|
    \frac{\partial \mathcal{L}}{\partial p_i^k}
    \right|}{\frac{\partial \mathcal{L}}{\partial p_i^k}}
\end{equation}
and has been recently used with neural network wave functions in the context of the Hubbard model.\cite{Luo2018}
Here $\alpha$ is a random number in the interval $(0,1)$ and $\eta$ sets the overall scale of the random step size.
The motivation for allowing the step size to be random is that it may help the optimization escape local minima that it encounters.
Within VMC, this algorithm can be run with fewer samples per iteration even compared to other gradient descent based algorithms as only the sign of the derivative needs to be known, but it typically requires many more iterations to converge.

ADAM and AMSGrad are popular methods within the machine learning community\cite{Ruder2016,Reddi2018,Kingma} and have similar forms. ADAM is given by:
\begin{equation}
\label{eqn:adamUpdate}
    p_i^{k+1} = p_i^{k} - \eta \frac{m_i^k}{\sqrt{n_i^k}}
\end{equation}
\begin{equation}
\label{eqn:adamMomentum}
    m_i^k = (1-\beta_1)m_i^{k-1} + \beta_1 \frac{\partial \mathcal{L}}{\partial p_i^k}
\end{equation}
\begin{equation}
\label{eqn:adamStep}
    n_i^k = \beta_2 \hspace{0.5mm} n_i^{k-1}
            + (1-\beta_2)\bigg(\frac{\partial \mathcal{L}}{\partial p_i^k}\bigg)^2
\end{equation}
AMSGrad is a recent adaptive step size scheme developed in response to the limitations of ADAM \cite{Reddi2018} and has almost the same form except for a slightly different denominator.
\begin{equation}
\label{eqn:amsgradStep}
    n_i^k = \mathrm{max}\Bigg(n_i^{k-1}, \hspace{2mm}
                (1-\beta_2) \hspace{0.5mm} n_i^{k-1} +
                \beta_2 \bigg(\frac{\partial \mathcal{L}}{\partial p_i^k}\bigg)^2\Bigg)
\end{equation}

In our calculations, we have used $\beta_1 = 0.1$ and $\beta_2 = 0.01$ for both AMSGrad and ADAM in line with the choice made by the Sharma group.\cite{Sabzevari2018,Mahajan2019}
It may be worth noting that a different convention appears in machine learning literature using $1-\beta_1$ and $1-\beta_2$ for what we and the Sharma group call $\beta_1$ and $\beta_2$. \cite{Ruder2016,Reddi2018}

Compared to the LM, these first derivative descent methods have some significant advantages.
Their low memory usage and reduced nonlinear bias make them a natural fit for the large parameter sets that the LM struggles to handle.
They are remarkably robust in the presence of noise and do not need special safeguards against statistical instabilities such as the LM's shifts.
At a basic practical level, the descent methods are also far simpler to implement than the LM and especially its blocked variant.
However, as we will see in our results, they often struggle
to reach the vicinity of the minimum using a comparable sampling effort.

\subsection{A Hybrid Optimization Method}
\begin{figure}
    \centering
    \includegraphics[width=8.3cm]{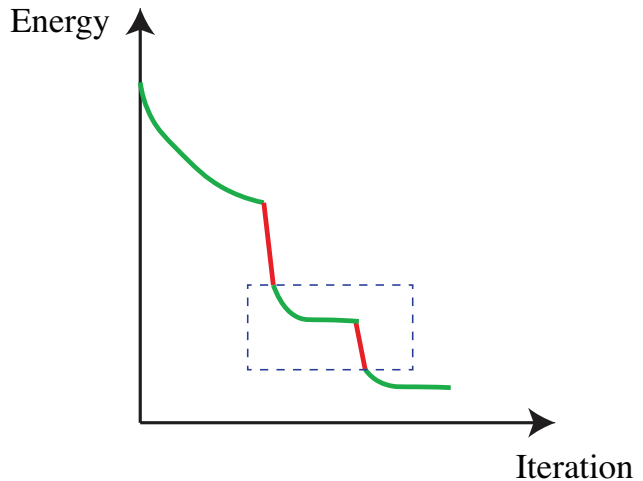}
    \caption{Schematic depiction of a typical energy optimization using the hybrid method. The dashed box around a section of descent in green and BLM in red defines a macro-iteration of the method.}
    \label{fig:hybridschematic}
\end{figure}

In an attempt to retain the benefits of both the LM and the AD
techniques, we have developed a hybrid optimization scheme that can be applied to large numbers of parameters.
Our approach alternates between periods of optimization using AD and sections using the BLM. 
Among other advantages, this allows us to use gradient descent to identify the $N_o$ previous important directions in parameter space that are used in the BLM via equation \ref{eqn:blmHistory}. 
The precise mixture of both methods can be flexibly altered, but a concrete example would be to first optimize for 100 iterations using RMSprop.
By storing a vector of parameter value differences every 20 iterations, we would produce 5 vectors that can be used for equation \ref{eqn:blmHistory} in some number (say three) steps of the BLM. 
After the execution of these BLM steps, the algorithm would return to another 100 iterations of descent and the process repeats until the minimum is reached.
Figure \ref{fig:hybridschematic} shows a generic depiction of how the ground state energy optimization may behave over the course of the hybrid method.
There are extended sections of computationally cheap optimization using gradient descent interwoven with substantial energy improvement over a few BLM steps.

\begin{figure}
    \centering
    \includegraphics[width=8.3cm]{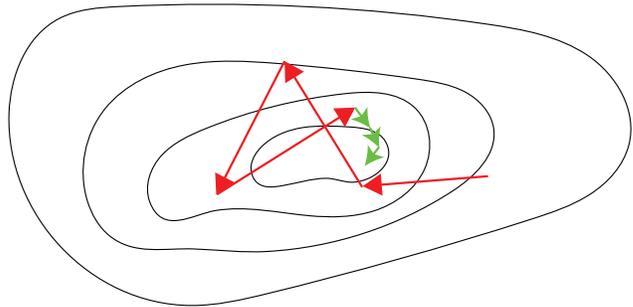}
    \caption{Schematic representation of gradient descent corrections in green to the red BLM steps, which we have observed to produce a greater degree of uncertainty about the location of the final minimum.}
    \label{fig:hybridcontour}
\end{figure}

The use of AD and the BLM should naturally allow parameter sets beyond the traditional LM limit of about 10,000 variables to be addressed, a limit we will surpass in the present study in the diflurodiazene system.
For now, Table \ref{tab:memCost} lays out how the memory cost of the methods we are considering scales with number of parameters $N$.
Both the hybrid method and the BLM steps it contains have a memory scaling that is intermediate between that of the standard LM and the descent methods.
The cost is given only approximately because while it is normally dominated by the cost of the $N_b$ blocks in the BLM, there are additional contributions related to how many directions are retained from the first BLM diagonalization and how many old directions are used.\cite{Zhao2017}

\begin{table}[htbp]
  \centering
  \small
  \caption{Rough memory cost scaling for the optimization methods we examine, with $N$ the number of optimized parameters and $N_b$ the number of blocks}
    \begin{tabular}{llll}
    Method Type & \multicolumn{1}{l}{Memory Cost}        \\ \hline
                & \\
    Standard Linear Method & $O(N^2)$ \\ 
               
                & \\
    Blocked Linear Method & $\thicksim O\left(\frac{N^2}{N_b}\right)$ \\ 
            
             & \\
    Hybrid Method & $\thicksim O\left(\frac{N^2}{N_b}\right)$ \\ 
             
              & \\
    Descent Methods & $O(N)$
      
    \end{tabular}%
  \label{tab:memCost}%
\end{table}%

One key motivation for including sections of AD, especially when the method is near convergence, is to counteract the noise we observe in LM updates.
While the LM tends to converge in a relatively small number of steps, we find the individual energies still fluctuate from iteration to iteration by multiple m$E_h$, particularly when we are working with wave functions that possess many highly nonlinear parameters.
Figure \ref{fig:hybridcontour} shows a cartoon of this behavior near the minimum
that prevents tight convergence.
Unless the shifts are large enough to constrain it to very small steps, the LM will tend to bounce around near the true minimum due to substantial (and biased) statistical uncertainties in its step direction.
The resulting energy fluctuations lead to ambiguity in what to report as the definitive LM energy.
One could take the absolute lowest energy reached on any iteration, but this is fairly unsatisfactory as it feels too dependent on a "lucky" step landing right on the minimum.
Our practice has been to take an average over multiple steps at the end of the optimization when parameter values should be converged.
However, this will generally include iterations with upward energy deviations due to the step uncertainties.
The use of AD offers a way out of this dilemma because it can correct the errors in the LM steps by moving towards the minimum more smoothly and with less bias.
As we shall demonstrate in our results, these considerations seem to give the hybrid method a statistical advantage over the LM by achieving lower error bars for the same computational cost.
They are also the basis of our recommendation for finishing optimizations with a long section of pure AD, which we shall show tends to improve the energy and greatly diminish the final statistical uncertainty.

\subsection{Wave Functions}

An assessment of optimization methods' effectiveness requires consideration of the form of the wave function that they are applied to. 
Multi-Slater determinant wave functions have been a common choice of ansatz in QMC and are typically combined with Jastrow factors that help recover some electron correlation and describe particle cusps.\cite{Foulkes2001} 
We specify our Multi-Slater Jastrow (MSJ) wave function with the following set of equations.

\begin{equation}
\label{eqn:psi}
    \Psi = \psi_{MS} \psi_J \psi_C
\end{equation}
\begin{equation}
\label{eqn:psiMS}
    \psi_{MS} = \sum_{i=0}^{N_D} c_i D_i
\end{equation}
\begin{equation}
\label{eqn:psiJ}
    \psi_J = \exp{\sum_i \sum_j \chi_k(|r_i - R_j|) + \sum_k \sum_{l>k} u_{kl} (|r_k - r_l|)}
\end{equation}

\begin{equation}
\label{eqn:psiNCJF}
    \psi_C = \exp(\sum_{IJ} F_{IJ} N_I N_J + \sum_K G_K N_K) 
\end{equation}

In equation $29$ above, $\psi_{MS}$ consists of $N_D$ Slater determinants $D_i$ with coefficients $c_i$.
It can be generated by some other quantum chemistry calculation such as complete active space self-consistent field (CASSCF) or a selective CI method prior to the VMC optimization.
In the one- and two-body Jastrow factor $\psi_J$, we have
functions $\chi_k$ and $u_{kl}$, which are constructed from
optimizable splines whose form is constrained so as to enforce
any relevant electron-electron and electron-nuclear cusp conditions.
\cite{Kim2018}

While MSJ wave functions with these types of traditional Jastrow
factors (TJFs) have been successfully used in many contexts,
\cite{Foulkes2001,Umrigar2007,Clark2011,Assaraf2017,Flores2019}
more involved correlation factors can be considered.
Typically, this involves the construction of many-body Jastrows
factors, \cite{Umrigar1988,Huang1997,Casula2003}
which may involve various polynomials of interparticle distances 
\cite{Huang1997,Luchow2015,LopezRios2012}
or an expansion in an atomic orbital basis
\cite{Casula2003,Casula2004,Beaudet2008,Sterpone2008,
      Marchi2009,Barborini2012,Zen2015}
or a set of local counting functions. \cite{Goetz2017,Goetz2018}
The latter case of many-body Jastrows, known as
real space number-counting Jastrow factors (NCJF),
is employed here as an example many-body Jastrow factor.
In real space, Jastrow factors have historically been effective
at encoding small changes to the wave function associated
with weak correlation effects, \cite{Foulkes2001}
but work in Hilbert space and lattice model VMC
reminds us that they can also be used to aid in the
recovery of strong correlations.
\cite{gutzwiller_gf,Neuscamman2013,Neuscamman2016}
One way to view NCJFs is as an attempt to develop a real space
many-body Jastrow factor that can aid in recovering both
strong and weak electron correlations.
\cite{Goetz2018}

The form of our NCJFs in equation \ref{eqn:psiNCJF} has the same structure as previously proposed four-body Jastrow factors,\cite{Marchi2009} where $N_I$ denotes the population of a region and the $F_{IJ}$ and $G_K$ are linear coefficients.
The region populations are computed by summing the values of counting functions at each electron coordinate.
\begin{equation}
\label{eqn:ncjfPop}
    N_I = \sum_i C_I (\mathbf{r_i})
\end{equation}
In this work, we use a recently introduced \cite{Goetz2018} form for the counting functions consisting of normalized Gaussians.
\begin{equation}
\label{eqn:ncjfCount}
    C_I = \frac{g_I(\mathbf{r})}{\sum_j g_j (\mathbf{r})}
\end{equation}
where 
\begin{equation}
\label{eqn:ncjfGauss}
    g_j(\mathbf{r}) = \exp ((\mathbf{r} - \mu)^T \mathbf{A} (\mathbf{r} - \mu) + K)
\end{equation}
describes a Gaussian about a center $\mu$. 
By placing these normalized Gaussians at various centers, we can divide up space with a Voronoi tessellation.
Schemes have been developed to generate partitions that either consist of regions centered on atoms or of finer grained divisions of space that can capture correlation within an atomic shell.
We make use of both types of partitioning methods for different wave functions in our study.
For simplicity, we only consider optimization of the parameters $F_{IJ}$ in the $F$-matrix of our NCJFs (the coefficients $G_K$ can be eliminated with a basis transformation of the region populations $N_I$),
\cite{Goetz2018}
but in principle the parameters defining the Gaussians $g_j$ could also be optimized.
We provide details of the Gaussians used in our ansatzes in Appendix D.

We also consider the problem of optimizing the molecular orbital
shapes alongside the other variational parameters. 
The ability to relax orbitals is important for successful study of many systems, particularly those involving excited state phenomena.\cite{Flores2019}
We make use of considerable theoretical and computational machinery based on the table method enhancements developed by Filippi and coworkers \cite{Filippi2016,Assaraf2017} that enables efficient evaluation of orbital rotation derivatives in large MSJ wave functions.
A rotation of molecular orbitals can be described with a unitary transformation with matrix $\mathbf{U}$ parameterized as the exponential of an antisymmetric matrix $\mathbf{X} = - \mathbf{X^T}$
\begin{equation}
\label{eqn:unitaryMat}
    \mathbf{U} = \exp (\mathbf{X})
\end{equation}
Impressively, one can obtain all wave function derivatives with
respect to the elements of $\mathbf{X}$ for a large multi-Slater
determinant ansatz for a cost that is only slightly higher than
that of the local energy evaluation.
For the details of how this is accomplished, we refer the reader
to the original publications. \cite{Filippi2016,Assaraf2017}
From the standpoint of parameter optimization, the main significance
of the orbitals (and the NCJFs) lies in both their nonlinearity
and  their strong coupling to other optimizable parameters.
In practice, we find that turning on the optimization of orbitals
and NCJFs greatly enhances the difficulty of the optimization problem
compared to MSJ optimizations in which only the CI 
coefficients and one- and two-body Jastrow parameters are varied.

\section{Results}

\subsection{Multi-Slater Jastrow N$_2$}
\label{sec:n2simple}

For a small initial test system, we consider the nitrogen dimer $\text{N}_2$ at the near-equilibrium and stretched bond lengths of 1.1 and 1.8 \r{A}. 
The nitrogen dimer is a known example of a strongly correlated system and a common testing ground for quantum chemistry methods.\cite{Langhoff1974,Rossi1999,Chan2004,Braida2011,Mazziotti2004,Neuscamman2013,Neuscamman2016}
The initial wave function ansatz consists of a modest number of Slater determinants (67 for the equilibrium geometry and 169 for the stretched, the result of a 0.01 cutoff limit on determinant coefficients) with traditional one-body and two-body Jastrow factors.
The Jastrow splines provide 30 additional optimizable parameters via 10 point cubic b-splines with cutoff distances of 10 bohr for the electron-nuclear and same-spin and opposite-spin electron-electron components.
The Slater determinant expansion is the result of a (10e,12o) CASSCF calculation in GAMESS\cite{Baldridge1993} using BFD pseudopotentials and the corresponding VTZ basis set.\cite{Burkatzki2007}
Due to the simplicity of the variable space in this case, we have
employed the $|\Psi|^2$ guiding function for all optimization
methods, including the LM and BLM.
See Appendix B for further computational details.

The first and simplest study we can make is to optimize our ansatzes with our multiple optimization techniques until convergence and compare final energies.
Note that all of our VMC optimizations with different methods in this study were performed using our implementations within a development version of the QMCPACK software package.\cite{Kim2018}
As N$_2$ is a small enough system that the traditional LM can be easily 
employed, we take the approach of first obtaining a traditional LM
optimization result and then using it as a reference against which
to compare the performance of other methods.
For the gradient descent methods, multiple optimizations were attempted 
with the initial step sizes tweaked from run to run based on a rough 
examination of how parameter values compared to the LM's results.
We find that the chosen values for the step sizes and other
hyperparameters in the gradient descent algorithms often
leads to apparent convergence at different energies.
It is therefore essential to make effective choices for these
parameters, which in part seems to rely on one's experience
with a given system.

\begin{figure}
    \centering
   \includegraphics[width=8.3cm]{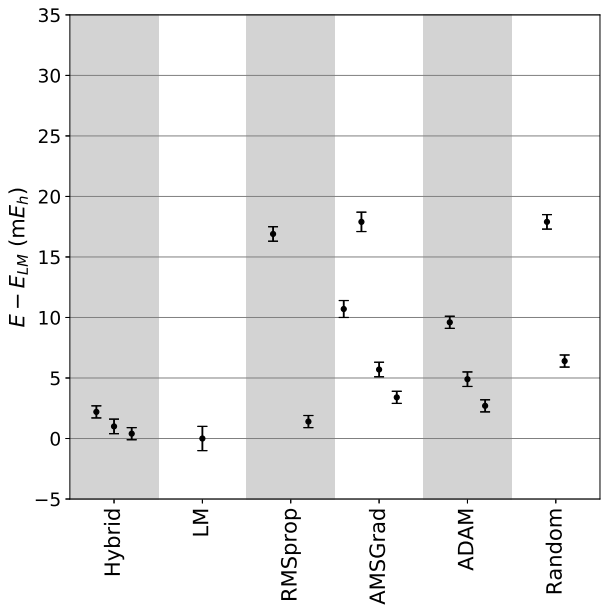}
    \caption{Different methods' optimized energies relative to that of the
         LM for equilibrium $\text{N}_2$ when optimizing CI coefficients
         and the TJF.
         }
    \label{fig:n2citjfeq}
\end{figure}

\begin{figure}
    \centering
   \includegraphics[width=8.3cm]{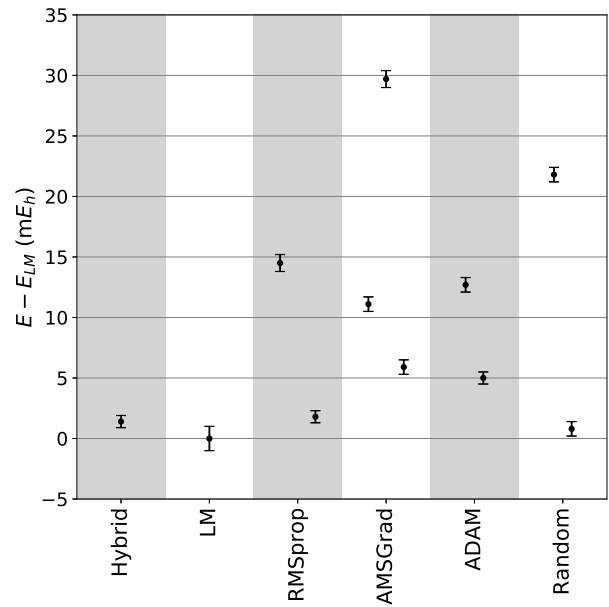}
    \caption{Different methods' optimized energies relative to that of the
         LM for stretched $\text{N}_2$ when optimizing CI coefficients
         and the TJF.
         }
    \label{fig:n2citjfst}
\end{figure}

Figures \ref{fig:n2citjfeq} and \ref{fig:n2citjfst} show energy differences relative to the LM result when optimizing the equilibrium and stretched nitrogen dimer wave functions respectively.
Tables providing the precise energies and statistical uncertainties as well as the step sizes used are shown in the appendices.
First, we see the choice of step sizes can have a substantial influence on the quality of gradient descent results.
In some cases, the same method can appear to converge to energies more
than 20 m$E_h$ apart when run with different initial step sizes.
While many of the gradient descent optimizations clearly did not reach
the minimum, the energy differences from the LM are only about 5 m$E_h$
or less when looking at the runs that used what turned out to be the
best choices for the hyperparameters.
With further tweaking of the hyperparameters, we would guess that at least
some of these descent methods could match the performance of the LM
in this simple test case.
Finally, we observe that the hybrid method performs about as well
as the best descent optimizations, typically reaching energies that
agree with the LM within error bars.

\subsection{All parameter N$_2$}
\label{sec:n2all}

We now add a NCJF and enable orbital optimization in order
to extend the comparison in a setting with a larger number
and variety of nonlinear parameters.
We will consider the relative merits of the optimization methods
in much greater detail in this setting as it offers
a clearer view of their differences.
For the number-counting Jastrow factor, we generated a set of 16 counting regions with 8 octants per atom after dividing space in half with a plane bisecting the bond axis.
The details are given in Appendix D, but we will note here that
this adds 135  $F$-matrix parameters to the optimization.
Allowing for orbital optimization adds another 663 and 618 parameters for the equilibrium and stretched cases, respectively.
Note that our implementation of orbital optimization in QMCPACK removes rotation parameters for orbitals that are not occupied in any determinant and also between orbitals occupied in all determinants, and so the precise number of rotation parameters is a function of the determinant expansion.
With orbital optimization enabled, the choice of importance sampling
function becomes an issue, and we now employ $|\Phi|^2$ 
for all LM and BLM steps with the $\epsilon$ weight set to 0.001.

\begin{figure}
    \centering
    \includegraphics[width=8.3cm]{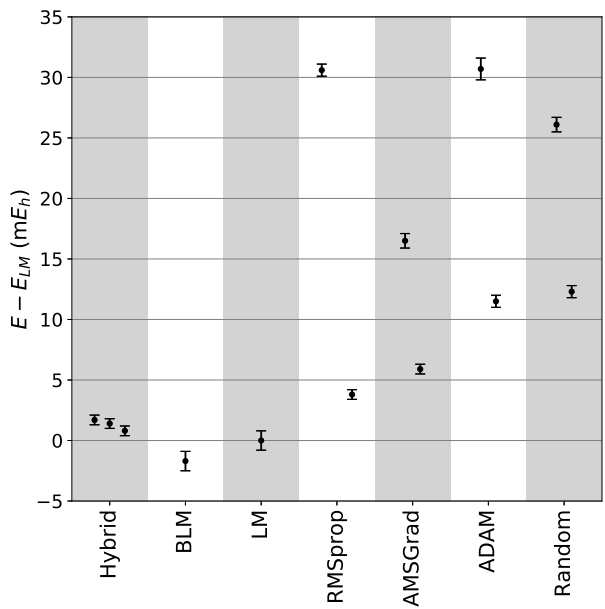}
\caption{Different methods' optimized energies relative to that of the
         traditional LM for equilibrium $\text{N}_2$ when all parameters
         are optimized simultaneously.
         See also Table \ref{tab:n2alleqData}.}
    \label{fig:n2alleq}
\end{figure}

\begin{figure}
    \centering
    \includegraphics[width=8.3cm]{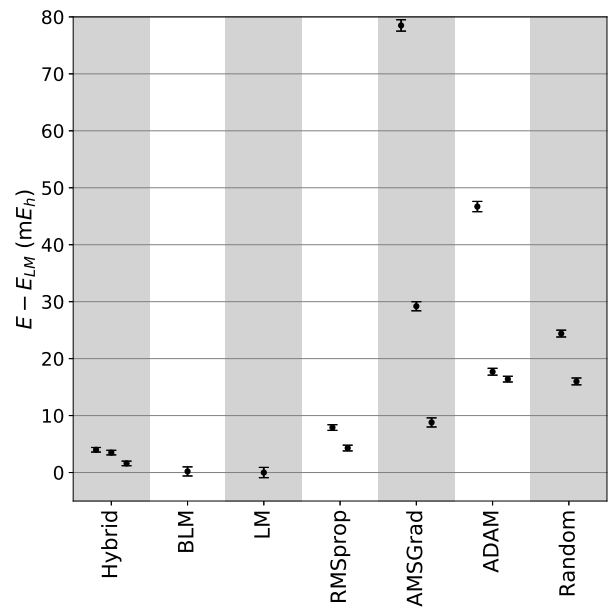}
    \caption{Different methods' optimized energies relative to that of the
         traditional LM for stretched $\text{N}_2$ when all parameters
         are optimized simultaneously.
         See also Table \ref{tab:n2allstData}.}
    \label{fig:n2allst}
\end{figure}

Figures \ref{fig:n2alleq} and \ref{fig:n2allst} show converged ground state energies relative to that of the LM.
For this more difficult version of the nitrogen dimer, we find that the gradient descent methods are less effective.
They now often yield energies that can be 10 m$E_h$ or more above the LM's answer though we again find that choice of step size plays a significant role.
The worst results for AMSGrad and ADAM were the result of choosing inappropriately large step sizes and simple reductions in the initial step size produced improvements in energy of tens of m$E_h$ though the final result still remained well above the LM's.
When we examined the optimizations over the course of their iterations, the gradient methods typically displayed some ability to quickly improve the wave function and energy initially, but they would then plateau and only very slowly improve the energy thereafter.
Extrapolating from our results indicates that even if these gradient descent methods eventually converge to the minimum, they will only do so after thousands more iterations and at a computational cost well beyond that of the LM.

\begin{table}[htbp]
  \centering
  \footnotesize
  \caption{Energies, uncertainties, and sample numbers for optimization of all parameters in equilibrium N$_2$.}
    \begin{tabular}{llll}
    Method & \multicolumn{1}{l}{Energy (a.u.)}        & \multicolumn{1}{l}{Uncertainty (a.u.)} & \multicolumn{1}{l}{Samples} \\ \hline
 
    Hybrid  1& -19.9263        & 0.0004      &5,400,000 \\ 
      Hybrid  2& -19.9266        & 0.0004        &10,000,000\\
         Hybrid  3& -19.9272        & 0.0004     &42,000,000\\ 
          &             &       &           \\\hline
    RMSprop 1& -19.8974        & 0.0005   & 20,000,000\\
    RMSprop 2& -19.9242        & 0.0004 & 20,000,000\\
          &       &       &  \\\hline
    AMSGrad 1& -19.9115      & 0.0006 & 20,000,000\\
        AMSGrad 2& -19.9221        & 0.0004 & 20,000,000\\
          &       &       &  \\\hline
    ADAM   1& -19.8973        & 0.0009 & 20,000,000\\
          ADAM   2& -19.9165        & 0.0005     & 20,000,000 \\
          &       &       &  \\\hline
    Random  1& -19.9019        & 0.0006 & 20,000,000\\
          Random  2& -19.9157     & 0.0005 & 20,000,000 \\
          &             &     & \\\hline
    LM    & -19.9280        & 0.0008     & 40,000,000\\
     &             &     & \\\hline
     BLM    & -19.9297        & 0.0008    & 80,000,000\\
      &             &     & \\\hline
      DF-BLM & -19.9293  &0.0001    & 90,000,000\\
      DF-Hybrid  1& -19.9290  &0.0001  & 15,400,000\\
      DF-Hybrid  2& -19.9290  &0.0001  & 20,000,000\\
      DF-Hybrid  3& -19.9293  &0.0001  & 52,000,000\\
      
    \end{tabular}%
  \label{tab:n2alleqData}%
\end{table}%

\begin{table}[htbp]
  \centering
  \footnotesize
  \caption{Energies, uncertainties, and sample numbers for optimization of all parameters in stretched N$_2$.}
    \begin{tabular}{llll}
    Method & \multicolumn{1}{l}{Energy (a.u.)}        & \multicolumn{1}{l}{Uncertainty (a.u.)} & \multicolumn{1}{l}{Samples}\\ \hline
       
    Hybrid  1 & -19.6316       & 0.0004    & 5,400,000\\
    Hybrid  2      & -19.6321        & 0.0004 & 8,400,000\\
     Hybrid  3     & -19.6340       & 0.0004 & 49,200,000\\
          &             &       &           \\ \hline
    RMSprop 1& -19.6277        & 0.0005      & 20,000,000\\
      RMSprop 2    & -19.6313        & 0.0005 & 20,000,000\\
          &       &       &  \\ \hline
    AMSGrad 1& -19.5571        & 0.0010 & 20,000,000\\
     AMSGrad 2    & -19.6064        & 0.0008 & 20,000,000\\
     AMSGrad 3    & -19.6268        & 0.0008 & 20,000,000\\
          &         &           &           \\ \hline
    ADAM 1& -19.5889        & 0.0009 & 20,000,000\\
     ADAM 2     & -19.6179        & 0.0006 & 20,000,000\\
      ADAM 3    & -19.6192        & 0.0005 & 20,000,000\\
          &         &           &           \\ \hline
    Random 1& -19.6112        & 0.0006 & 20,000,000\\ 
     Random 2       &-19.6196        & 0.0006 & 20,000,000\\
          &              &       & \\ \hline
    LM    & -19.6356        & 0.0009 & 40,000,000\\
     &              &       & \\ \hline
     BLM    & -19.6354        &0.0008 & 80,000,000\\
     &              &       & \\ \hline
     DF-BLM & -19.6356  &0.0001 & 90,000,000\\
      DF-Hybrid 1& -19.6352  &0.0001 & 15,400,000\\
      DF-Hybrid 2& -19.6354  &0.0001 & 18,400,000\\
      DF-Hybrid 3& -19.6346  &0.0001 & 59,200,000\\
      
    \end{tabular}%
  \label{tab:n2allstData}%
\end{table}%

\begin{figure}
    \centering
    \includegraphics[width=8.3cm]{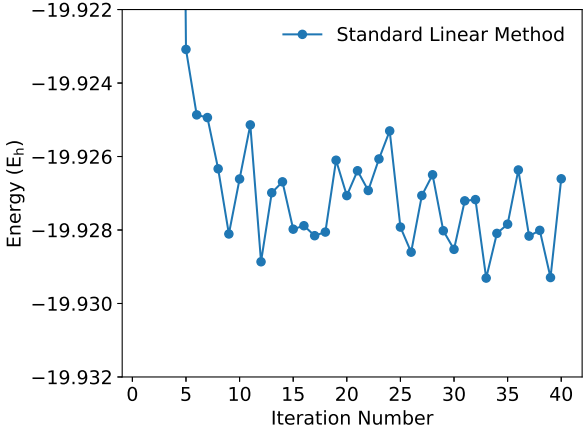}
    \caption{The standard LM optimization for all parameter equilibrium N$_2$.}
    \label{fig:n2lm}
\end{figure}

A more careful comparison of the different methods can be made by referring to Tables \ref{tab:n2alleqData} and \ref{tab:n2allstData}, which list the precise converged energies and their error bars.
We also report the total number of samples used in each optimization as a proxy for computational effort, noting that for the BLM and the BLM portion of the hybrid method we double counted samples out of fairness as the BLM steps require running over their samples twice.
In assessing cost, one must also consider the statistical uncertainty achieved, where we see that the LM and BLM are at a disadvantage.
To help illustrate the update uncertainty contribution to this error, which we first discussed in the theoretical section above, we show the energy versus LM iteration for equilibrium N$_2$ in Figure \ref{fig:n2lm}.
The fluctuations in energy from step to step, sometimes by as much as 2 m$E_h$, demonstrate the difficulty the LM faces from the uncertainty in its steps near the minimum.
In this case, we see that the LM's final energy uncertainty is driven by the update step uncertainty rather than the uncertainty in evaluating the energy for a given set of parameter values at a particular iteration.
We have observed similar behavior in the BLM and include the result of a BLM calculation in the tables.
Note that the tabulated energies come from an average over the last ten optimization steps in the case of the standard LM and BLM and from an average over the last 50 descent steps in the case of the hybrid and pure descent methods.

\begin{figure}
    \centering
    \includegraphics[width=8.8cm]{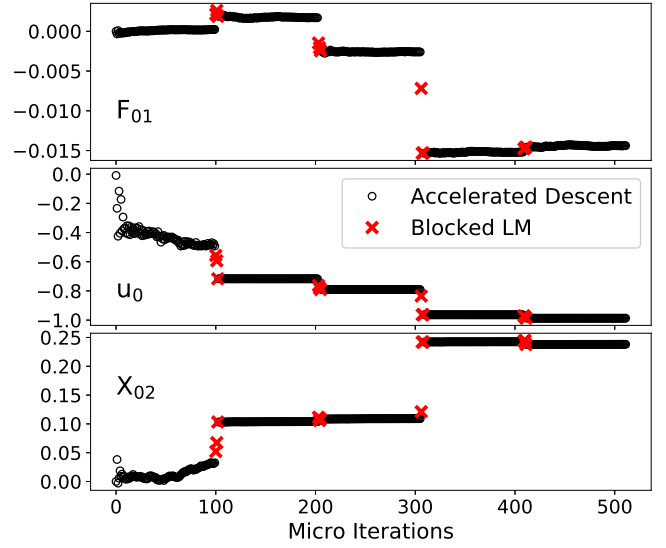}
    \caption{
    Values for the first off-diagonal $F$-matrix element ($F_{01}$),
    the first electron-nuclear TJF spline parameter ($u_0$),
    and the second orbital rotation variable ($X_{02}$) at each
    micro iteration of the ``Hybrid 1'' optimization for
    all parameter equilibrium N$_2$.
    }
    \label{fig:n2param}
\end{figure}

From the tabulated data, we see that the hybrid optimization can achieve
lower energies than the gradient descent methods using fewer samples,
and that its results are typically within a few m$E_h$ of the traditional LM.
While the accelerated descent sections of the hybrid method provide
some swift energy reductions early on when the wave function is still far
from the minimum in parameter space, the BLM steps in the algorithm
greatly accelerate the process of bringing the parameters near to the
minimum, as can be seen in Figure \ref{fig:n2param}.
Looking at the electron-nuclear spline parameter and the orbital
rotation parameter, we see typical
cases in which rapid initial parameter movement during the early
part of the first RMSprop stage transitions to much slower movement
later in that stage, followed by very little movement at all
in later RMSprop stages.
Note that the latter can be explained largely by the need to
keep initial step sizes small in later stages to avoid
significant upward deviations in the energy as the RMSprop method
rebuilds its momentum history.
In between these AD stages, the BLM updates move the parameter
values in much larger steps, greatly accelerating convergence.
This behavior makes the hybrid approach somewhat more black box
as compared to the pure descent approaches, as the ability to
get near the minimum with a modest sampling effort is much
less dependent on the choice of the initial step sizes than for
the AD methods.
This conclusion is supported by the fact that the hybrid
optimizations in Tables \ref{tab:n2alleqData} and \ref{tab:n2allstData}
used various initial step size settings (as discussed in Appendix B)
and nonetheless  produced lower energies than the pure descent methods
in every case.

\begin{figure}
    \centering
    \includegraphics[width=8.3cm]{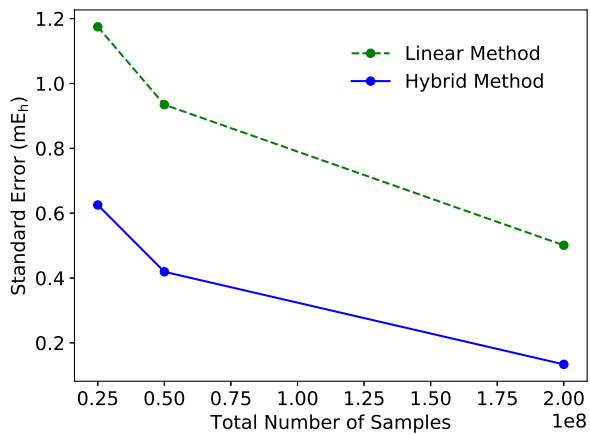}
    \caption{Standard Errors for the hybrid method and LM on all parameter
    equilibrium $\text{N}_2$ vs different optimizations'
    total sampling costs.}
    \label{fig:n2eqstderr}
\end{figure}

\begin{figure}
    \centering
   \includegraphics[width=8.3cm]{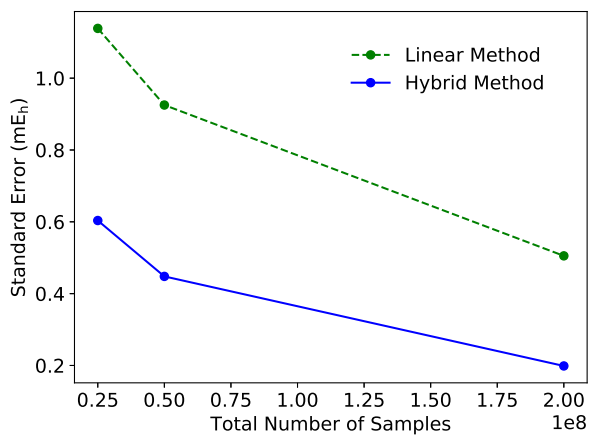}
    \caption{Standard Errors for the hybrid method and LM on
    all parameter stretched $\text{N}_2$ plotted against different optimizations'
    total sampling costs.}
    \label{fig:n2ststderr}
\end{figure}

As discussed in our introduction of the hybrid method, another advantage
is its ability to obtain a lower error bar at convergence
than the LM for the same overall computational cost.
This is a natural consequence of spending part of its sampling effort
on gradient descent steps that correct for the BLM steps' uncertainty
and bias (as illustrated earlier in Figure \ref{fig:hybridcontour})
and that hew closer to
the zero-variance principle by importance sampling with $|\Psi|^2$.
To demonstrate this advantage explicitly, we ran additional sets
of LM and hybrid optimizations adjusted to have essentially the same
total number of samples.
We then compare the standard error for the last ten LM steps
and the last ten hybrid macro iterations in 
Figures \ref{fig:n2eqstderr} and \ref{fig:n2ststderr},
where we find that the hybrid has a substantially lower
statistical uncertainty in every case.
Assuming the usual $N^{-1/2}$ decay of uncertainty
with sample size, the LM would require a factor of roughly
four times more samples to reach the hybrid's uncertainty,


\begin{figure}
    \centering
    \includegraphics[width=8.3cm]{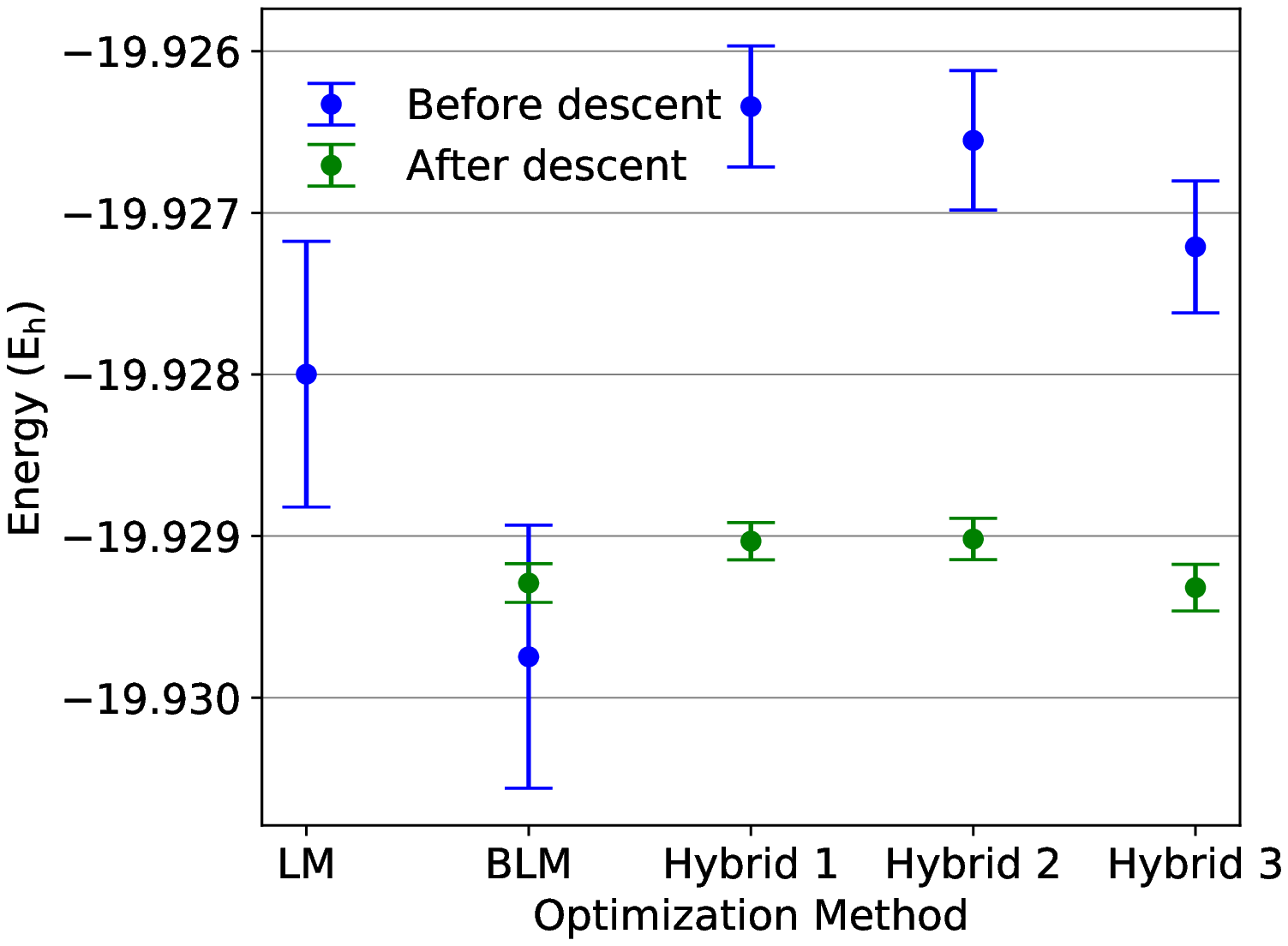}
    \caption{Converged energies in equilibrium $\text{N}_2$ before and after a final descent optimization.}
    \label{fig:n2dfeq}
\end{figure}

\begin{figure}
    \centering
    \includegraphics[width=8.3cm]{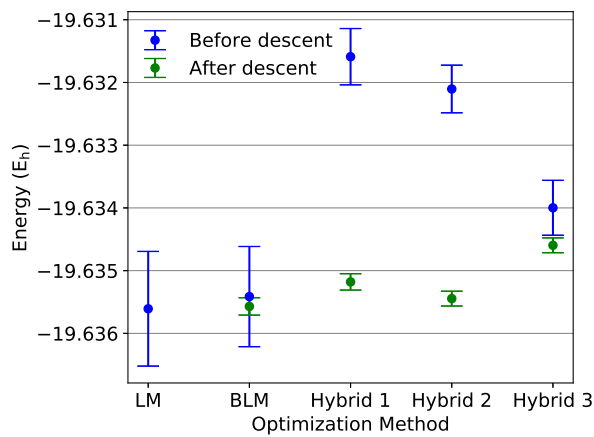}
    \caption{Converged energies in stretched $\text{N}_2$ before and after a final descent optimization.}
    \label{fig:n2dfst}
\end{figure}

These statistical advantages in the final energy can be improved even further if we finish an optimization with a long section of pure descent.
To demonstrate this, we have taken the final wave functions produced by the hybrid and BLM optimizations in Tables \ref{tab:n2alleqData} and \ref{tab:n2allstData} and applied a further period of optimization using RMSprop with initial step sizes of 0.001 for all parameters.
This ``descent finishing'' (DF) adds only a modest additional cost compared to the preceding optimization and yields a large improvement in statistical uncertainty and, in many cases, an improvement in the final energy value as well.
These advantages can be seen clearly in Figures \ref{fig:n2dfeq} and \ref{fig:n2dfst}, as well as in Tables \ref{tab:n2alleqData} and \ref{tab:n2allstData},
where we observe final error bars that are a factor of
eight smaller than those of the LM.
In terms of cost, this implies that the traditional LM would
have required 64 times the original number of samples
to achieve the DF-BLM or DF-hybrid precision.
Put another way, we find that the DF-hybrid
approach gives an equivalent or lower energy, with a much
smaller error bar, at a substantially lower cost.
Note that, in contrast, we find that this DF approach is not very effective
when used in conjunction with the pure descent methods, where
it essentially amounts to restarting the methods at the
parameter values found after the first run of their optimization.
While we do find that this restarting of the accumulation
of momentum can improve the energy, the wave function
parameters still do not reach their optimal values and the
energy lowering vs total sampling cost is not competitive
with the DF-hybrid.

Our study of the nitrogen dimer provides some clarity on the relative strengths of the LM and gradient descent, while also pointing the way to a more effective synthesis of the two.
Gradient descent methods struggle in the presence of a variety of different highly nonlinear parameters, although they did perform better when we were only optimizing TJFs and CI coefficients.
Among the descent methods, we found that the RMSprop approach came the
closest to achieving the LM minimum energy.
It is of course difficult to rule out the possibility that this
and other AD methods could reach the LM energy
with additional sampling and more experimentation with
the hyperparameters.
However, it is far from obvious that this would be
be cost-competitive, and the need to make careful and possibly
system-specific choices for hyperparameters
is somewhat antithetical to the general aspiration that an
optimizer be as black-box as possible.
For its part, the LM is more effective at moving parameters
into the vicinity of the minimum, but tight convergence is then stymied by
an unsatisfactory level of biased statistical uncertainty.
As a side note, this behavior --- in which the first derivative methods give
better convergence once near the minimum but are at a relative disadvantage
far from the minimum --- is somewhat the reverse of what one would expect
in deterministic optimization, where second derivative methods are at their
strongest relative to first derivative methods during the final tight
convergence in the vicinity of the minimum.
\begin{figure}[t]
    \centering
    \includegraphics[width=8.3cm]{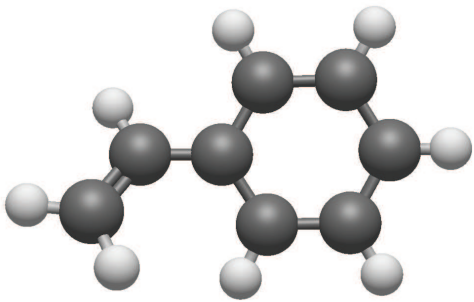}
    \caption{Equilibrium geometry of styrene. See Appendix C for structure coordinates.}
    \label{fig:styrene_geom}
\end{figure}
Although things are reversed in the stochastic VMC case, we stress that the
two classes of methodology are strongly complementary, as they compensate
for each other's weaknesses.
By using a low-memory version (BLM or hybrid) of the LM to get near
to the minimum and then handing off to an accelerated descent method
to achieve tight convergence, we find better overall performance than when
working with either class of method on its own.
These insights in hand, we will now apply this combined approach in a pair of larger and more challenging VMC optimization examples.

\subsection{Styrene}
We first turn to styrene at its equilibrium geometry 
(Figure \ref{fig:styrene_geom}) which offers an optimization with both
more electrons and more variables, but in which the traditional LM
is still quite achievable for comparison.
As in N$_2$, we construct a multi-Slater wave function modified by both TJFs and a NCJF.
To generate our Slater determinants, we have employed the
heatbath selective CI (HCI) method as implemented in the Dice
code by Sharma and coworkers. \cite{Holmes2016,Sharma2017}
The orbital basis for the HCI calculation was produced via a (14e,14o) CASSCF calculation in Molpro \cite{MOLPRO_brief} using a recently developed set of pseudopotentials and their corresponding double zeta basis.\cite{Bennett2017}
In this CASSCF basis, HCI then correlated 32 electrons (out of a total of 40 electrons left over after applying pseudopotentials) in 64 orbitals.
For our NCJF, we defined one counting region per atom, giving our $F$-matrix 135 optimizable parameters (see Appendix D for further NCJF details).

\begin{figure}[t]
    \centering
    \includegraphics[width=8.3cm]{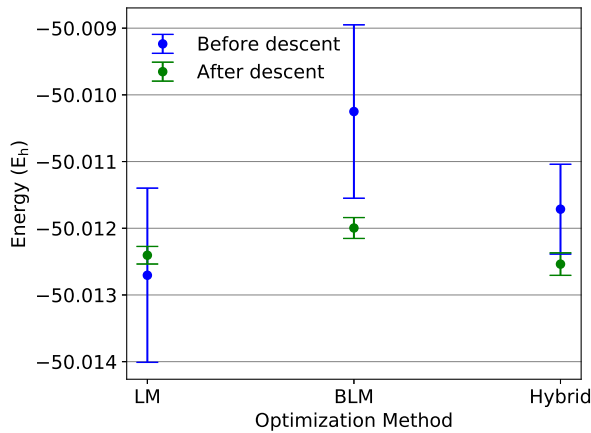}
    \caption{Converged energies in equilibrium styrene before and after a final descent optimization.}
    \label{fig:styrene_energies}
\end{figure}

We optimized our wave function in a staged fashion using the standard LM, the BLM, and the hybrid method.
First, we conducted a partial optimization of the TJFs and the 100 most important CI coefficients.
We then turned on the optimization of the orbitals and the NCJF's $F$-matrix, reaching a total of 4,570 parameters, most of them highly nonlinear.
In the hybrid and BLM optimizations, the parameters were divided into 5 blocks and we used $N_k = 50$ and $N_o = 5$ for our numbers of kept directions and previous important directions, respectively.
We used a value of 0.001 for $\epsilon$ in the $|\Phi|^2$ distribution for the LM and BLM sampling.
These optimizations were then followed by 1,000 steps of RMSprop.
As shown in Figure \ref{fig:styrene_energies}, we find that our hybrid method reaches a converged energy as low or better than that of the standard and blocked LM, and finishing our optimizations with descent provides a substantial improvement in the statistical uncertainty even in this more challenging case.

\begingroup
\setlength{\tabcolsep}{1.5pt}
\renewcommand{\arraystretch}{1.25} 
\begin{table}
  \centering
  \small
  \caption{A summary of the VMC optimization stages in FNNF showing
           the number of determinants
           $N_d$ included from HCI, which parameters are optimized,
           and the total number $N_p$ of optimized parameters.
           Note that CI coefficients are optimized at every stage.
           Stages 2, 3, and 4 start from the parameter values from
           the previous stage, with newly added determinants' coefficients
           initialized to zero.
           We also report the number of iterations performed in each stage,
           which for stage 4 is simply the number of RMSprop steps.
           A hybrid iteration, on the other hand, consists of 100 RMSprop steps
           followed by three BLM steps.
           All RMSprop steps use 20,000 samples drawn from $|\Psi|^2$,
           while the BLM steps each use 1 million samples drawn from the
           $|\Phi|^2$ guiding function with $\epsilon$ set to 0.01.
          }
    \begin{tabular}{cclcccrr}
    Stage & Method & $N_d$ & TJF & $F$-matrix & Orbitals & $N_p$ & Iterations \\
    \hline
    1 & Hybrid & $10^2$ & \checkmark & & & 139 & 9 \\
    2 & Hybrid & $10^3$ & \checkmark & \checkmark & & 1048 & 4 \\
    3 & Hybrid & $10^4$ & \checkmark & \checkmark & \checkmark & 15,573 & 6 \\
    4 & AD     & $10^4$ & \checkmark & \checkmark & \checkmark & 15,573 & 1,000 \\
    \end{tabular}%
  \label{tab:fnnf_stages}%
\end{table}%
\endgroup

\subsection{FNNF}
\label{sec:fnnf}

We now turn our attention to a strongly correlated transition
state of the the diflurodiazene (FNNF) \textit{cis-trans}
isomerization, where we test the hybrid optimization approach
on a much larger determinant expansion.
The FNNF isomerization  can be thought of as a toy model molecule
for larger systems such as photoswitches,
which have potential uses in molecular machines \cite{Russew2010,Kinbara2005}
and high-density memory storage. \cite{tian2004switches}
In addition, FNNF itself is of interest as part of the synthesis
of high energy polynitrogen compounds and has been the subject of
multiple electronic structure studies.\cite{Christe1991,Christe2010,Lee1989}
Here we focus on its strongly correlated transition state, which is
the direct analogue of the out-of-plane TS1 transition state 
in diazene. \cite{Sand2012}

\begin{figure*}
    \centering
    \includegraphics[width=18.5cm]{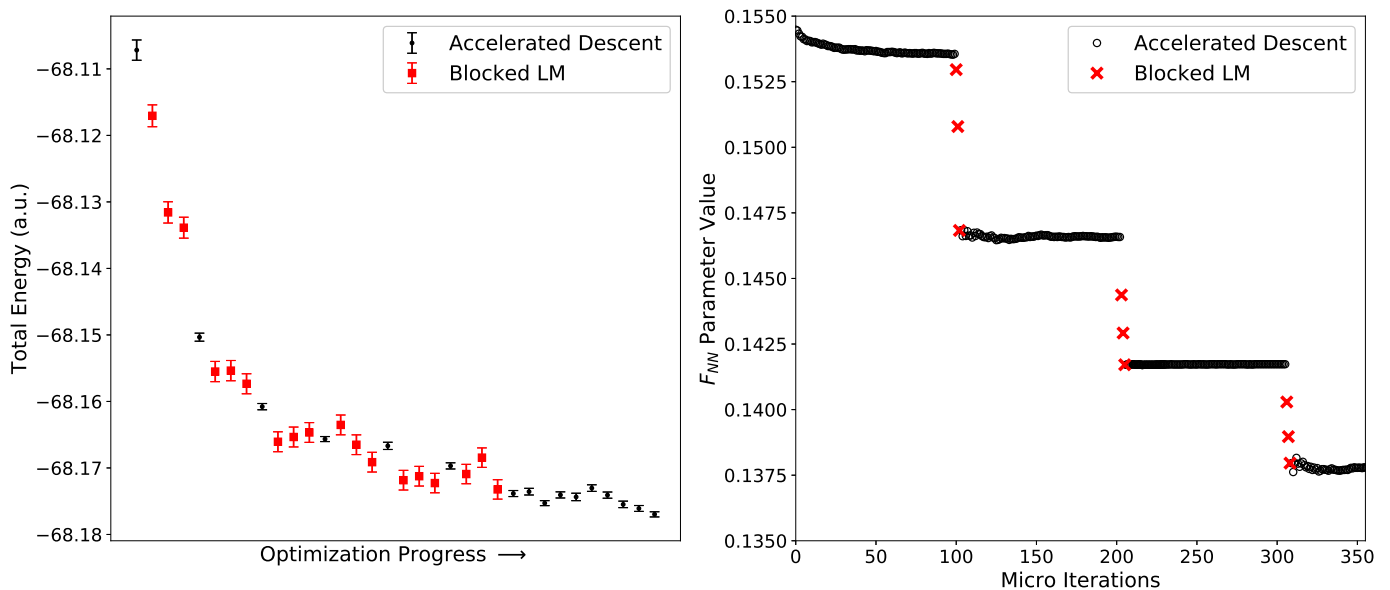}
    \caption{Left panel: energies during stages 3 and 4 of the FNNF
             optimization.  The descent energies are reported as the
             average over the last 50 RMSprop steps within each block
             of 100 RMSprop steps, whereas the BLM energies are
             the energy estimates on the random samples used for the
             BLM update steps.
             Right panel:  change in the value of the $F$-matrix
             parameter that couples the two nitrogen atoms' counting
             regions over the first three macro iterations
             of stage 3, with each micro iteration corresponding to
             one RMSprop or BLM step.
             The nine BLM points on the right panel correspond to
             the first nine BLM points on the left panel.
            }
    \label{fig:fnnf_plot}
\end{figure*}

Our treatment of this transition state began by locating its
geometry via an (8e,8o) CASSCF optimization in the cc-pVTZ basis
using Molpro.\cite{MOLPRO_brief}
At this geometry (given in Appendix C)
we then switch over to using BFD pseudopotentials
and their corresponding triple zeta basis,
\cite{Burkatzki2007}
in which we use the Dice code
\cite{Holmes2016,Sharma2017}
to iterate an HCI calculation
with 24 electrons
distributed in the lowest 50 (8e,8o) CASSCF
orbitals until its variational wave function
has reached almost 2 million determinants.
We then import the first 10,000 of these determinants
into our VMC optimization and combine them with
TJFs, atom-centered NCJFs, and orbital optimization,
which produces an ansatz with over 15,000 variational parameters.

Our VMC optimization proceeds in stages as summarized in
Table \ref{tab:fnnf_stages}.
This begins with TJFs and a 100-determinant ansatz from HCI,
with later optimization stages adding
more determinants and turning on the optimization of
the NCJF and orbital rotation variables.
As in styrene, the strategy is to bring the parameters
near to their optimal values with the help of the LM
and then to perform a final
unbiased relaxation via a long run of RMSprop AD.
Due to the large number of variational parameters,
we incorporate the LM via the hybrid scheme, with the BLM
steps employing 2, 2, and 10 blocks
during stages 1, 2, and 3, respectively.
In stages 1 and 2, we used an initial RMSprop step size of 0.01 for TJFs and CI coefficients
before setting it to 0.005 at the beginning of stage 3.
For the $F$-matrix parameters, we began by setting the initial
step size to 0.001,
but after observing a significant rise and fall of the energy
during the RMSprop section of the first hybrid macro iteration
in stage 3, we reduced this to 0.0001 and also lowered the TJF and CI step size to 0.0005
for the last 4 macro iterations in that stage.
For all steps in stage 4, we maintained the 0.0001 step size for the $F$-matrix parameters and lowered the initial step size for TJFs and CI coefficients to 0.0002.
An initial step size of 0.0001 was used for orbital parameters throughout both stages 3 and 4.
The BLM steps used the $|\Phi|^2$ guiding function with a value of 0.01 for $\epsilon$.

\begin{table}
  \centering
  \small
  \caption{Energies of the transition state of FNNF.}
    \begin{tabular}{llll}
    Method & \multicolumn{1}{l}{Energy (a.u.)} &       & \multicolumn{1}{l}{Uncertainty (a.u.)} \\
    \hline
   Hartree-Fock &  -67.112730 & &  \vphantom{\big(\big)} \\
    CASSCF & -67.359100 & &  \vphantom{\big(\big)} \\
       VMC Stage 1 &  -68.1017 & & 0.0011 \vphantom{\big(\big)} \\
       VMC Stage 2 & -68.1213 &  &0.0009 \vphantom{\big(\big)} \\
    VMC Stage 3 & -68.1698 &       & 0.0006 \vphantom{\big(\big)} \\
      VMC Stage 4 \hspace{2mm} & -68.1750 & &0.0002 \vphantom{\big(\big)} \\
    \end{tabular}%
  \label{tab:fnnf_energies}%
\end{table}%

The energies resulting from this staged optimization
are shown in Table \ref{tab:fnnf_energies} and Figure \ref{fig:fnnf_plot}.
Unsurprisingly, stage 3 proved to be the most challenging and expensive
stage, as it is where we hope to move all parameters near to their final
values in a setting where the traditional LM would face severe memory
bottlenecks.
As seen in Figure \ref{fig:fnnf_plot}, both the AD and BLM steps
clearly work to lower the energy during the first two macro iterations of
stage 3.
In the last four macro iterations of stage 3, however, the energy decreases
more slowly and it is less clear, at least when looking at the energetics,
whether the BLM steps are still necessary.
Instead, their importance is revealed by inspecting the movement of
the $F$-matrix values within the NCJF, an example of which is shown
in the right-hand panel of Figure \ref{fig:fnnf_plot}.
As in N$_2$, these parameters prove to be the most resistant to optimization
via AD, and we clearly see that although AD does gradually move their values
in the same direction as the BLM, the BLM steps dramatically accelerate
their optimization.
This effect is seen throughout all six macro iterations of stage 3, and
so although the BLM energies are not obviously improving at the end of this
stage, the inclusion of these steps is clearly still beneficial.
Note that relaxing the NCJF after moving from a 1,000-determinant to a
10,000-determinant expansion is important, because the larger determinant
expansion is better able to capture some of the correlation effects
that the NCJF is encoding, and so we expect (and indeed see) that this
diminishing of its role leads to smaller $F$-matrix values being optimal.

Although we have again found that it would be difficult for AD alone
to provide a successful optimization of our ansatz, the statistical
advantages of its incorporation are still quite clear.
A close inspection of the sample sizes used in the optimization
reveals that each of the AD and BLM points in the left panel of
Figure \ref{fig:fnnf_plot} corresponds to averaging over 1 million
random samples.
Despite this equal sampling effort, the uncertainties for the
AD energy estimates are about one third the size of those for the
BLM, implying that a pure BLM approach would require an order of
magnitude more sampling effort to produce similar results.
To understand this statistical advantage, we need to remember
two important differences between the AD and BLM steps.
First, the nonlinearity of the LM and BLM eigenvalue problem
leads to biases in the update steps that can both increase the
step-to-step energy uncertainty and cause the method to optimize
off-center from the true minimum.
Second, the use of an alternative guiding function for the
BLM samples in order to mitigate this step uncertainty moves
us away from the zero-variance regime enjoyed by traditional
$|\Psi|^2$ sampling.
If we were to instead employ traditional sampling, our energy
estimates for a specific wave function would improve,
but the BLM step uncertainty would increase sharply.
As the AD methods do not suffer from these issues, they help us to
further mitigate the BLM step uncertainty and to perform
a final, high-precision relaxation during stage 4.
In total, incorporating the AD steps in this case roughly
doubles the number of samples required, but is well worthwhile given
that it improves statistical efficiency by almost an order of magnitude.

While it is possible that the NCJF parameters are not quite converged
in this particular optimization
and that increasing iteration counts in stages 1 through 3 could
further improve the energy, the lessons learned from investigating
a large MSJ optimization for the FNNF transition state are already clear.
While both the BLM and the AD methods can be used in this 10,000+ parameter
regime, they bring highly complementary advantages to the optimization
and so would appear to work better together than apart.
In particular, the BLM helps optimize the parameters that change only
very slowly during AD, whereas the statistical advantages of AD
greatly increase precision at a given sample size and work to
eliminate the statistical biases suffered by the BLM.

\section{Conclusions and Outlook}
\label{sec:conclusion}

We have found that a combination of first and second derivative
optimization methods appears to work better than using either
class of method on its own when minimizing the energies of
wave functions in variational Monte Carlo.
This is particularly true for wave functions with a wide
variety of different types of nonlinear parameters, as for
example when dealing simultaneously with traditional
one- and two-body Jastrow factors, many-body Jastrow factors,
and orbital relaxations.
While the linear method and its low-memory variants show a
superior ability to move these nonlinear parameters into
the vicinity of their optimal values, accelerated descent
methods prove much more capable of converging them tightly
around the minimum.
This situation stands as an interesting reverse of what
is typically encountered in deterministic optimization,
where second derivative methods are usually superior
for tight final convergence and first derivative methods
perform relatively better in the early stages of an
optimization.
The realities of working with statistically uncertain
energies and energy derivatives turns this expectation
on its head, both because of the need to stabilize the
statistics of the linear method's second derivative
elements through zero-variance-violating importance
sampling schemes and due to the nonlinear biases that
are induced when solving the linear method's eigenvalue
problem.
The linear method's ability to quickly move the parameters
near the minimum, however, makes it appear that employing
it as part of a hybrid approach is well worthwhile.
Indeed, in our testing, hybridizing low-memory
linear method variants with accelerated descent methods
provides better energies with smaller
statistical uncertainties at a lower computational cost
when compared to the stand-alone use of either the linear method
or accelerated descent methods.

Looking forward, there are many questions still to be answered
about the interplay between first and second derivative methods.
For example, although the blocked linear method greatly reduces
memory cost vs the traditional linear method, it is not clear that
it can be applied effectively beyond 100,000 parameters in its
current form.
One thus wonders whether it is necessary to optimize all of the
parameters during the linear method steps of the hybrid
approach, or whether it may be possible to identify
(perhaps during an ongoing optimization?) which parameters
would benefit from linear method treatment and which would not.
Were such a sorting possible, accelerated descent methods with
their even lower memory footprint could be left to deal with
most of the parameters, with only a relatively small subset
treated by the linear method steps.
Another important issue is making the hybrid approach as black
box and user friendly as possible.
Although we have tested it here with many different descent
step size settings for the different parameter types, this has
not been a systematic survey.
More extensive testing may allow clear defaults to be settled
upon so that users can reasonably expect a successful
optimization without resorting to careful step size control.
We look forward to investigating these exciting possibilities
in future.


\section{Acknowledgements}
This work was supported by the Office of Science,
Office of Basic Energy Sciences, the US Department of Energy,
Contract No.\ {DE-AC02-05CH11231}.
Calculations were performed using the Berkeley Research Computing
Savio cluster and
the National Energy Research Scientific Computing Center,
a DOE Office of Science User Facility supported by the
Office of Science of the U.S. Department of Energy
under Contract No.\ {DE-AC02-05CH11231}.

\section{Appendix A: Additional Energies}

\begin{table}[H]
  \centering
  \footnotesize
  \caption{Precise Values for optimizing CI coefficients and traditional Jastrow factors in equilibrium N$_2$.}
    \begin{tabular}{llll}
    Method & \multicolumn{1}{l}{Energy (a.u.)}        & \multicolumn{1}{l}{Uncertainty (a.u.)} & \multicolumn{1}{l}{Samples}\\ \hline
    
    Hybrid 1& -19.9083        & 0.0005    & 3,900,000\\
     Hybrid 2     & -19.9095        & 0.0006  & 5,400,000\\
     Hybrid 3     & -19.9101        & 0.0005      & 29,000,000\\
          &         &           &           \\ \hline
    RMSprop 1& -19.8936        & 0.0006   & 20,000,000\\
    RMSprop 2& -19.9091        & 0.0005 & 20,000,000\\
          &       &       &  \\\hline
    AMSGrad 1& -19.8998        & 0.0007 & 20,000,000\\
    AMSGrad 2      & -19.8926        & 0.0008 & 20,000,000\\
    AMSGrad 3      & -19.9048        & 0.0006 & 20,000,000\\
     AMSGrad 4     & -19.9071        & 0.0005 & 20,000,000\\
          &             &       & \\ \hline
    ADAM 1& -19.9009        & 0.0005 & 20,000,000\\
     ADAM 2     & -19.9056        & 0.0006 & 20,000,000\\
     ADAM 3     & -19.9078        & 0.0005 & 20,000,000\\
          &         &           &           \\ \hline
    Random 1& -19.8926        & 0.0006 & 20,000,000\\
     Random 2   &-19.9041        &0.0005 & 20,000,000 \\
          &       &        & \\\hline
    Linear Method    & -19.9105        & 0.0010 & 40,000,000 \\
    \end{tabular}%
  \label{tab:n2tjfcieqData}%
\end{table}%

\begin{table}[H]
\footnotesize
  \centering
  \caption{Precise Values for optimizing CI coefficients and traditional Jastrow factors in stretched N$_2$.}
    \begin{tabular}{llll}
    Method & \multicolumn{1}{l}{Energy (a.u.)}        & \multicolumn{1}{l}{Uncertainty (a.u.)} & \multicolumn{1}{l}{Samples}\\ \hline
   
    Hybrid 1& -19.6141        & 0.0005    & 27,000,0000\\ 
          &       &       &  \\ \hline
    RMSprop 1& -19.6010        & 0.0007 & 20,000,000\\
     RMSprop 2& -19.6137        & 0.0005 & 20,000,000\\
          &       &       &  \\ \hline
    AMSGrad 1& -19.6044        & 0.0006 & 20,000,000\\
    AMSGrad 2      & -19.5858        & 0.0007 & 20,000,000\\
     AMSGrad 3     & -19.6096        & 0.0006 & 20,000,000\\
          &              &      &  \\\hline
    ADAM 1& -19.6028        & 0.0006 & 20,000,000\\
    ADAM 2      & -19.6105        & 0.0005 & 20,000,000\\
          &       &       &  \\ \hline
    Random 1& -19.5937        & 0.0006 & 20,000,000\\
    Random 2        &-19.6147        &0.0006 & 20,000,000 \\
          &             &        &  \\ \hline
    Linear Method    & -19.6155        & 0.0010 & 40,000,000 \\
    \end{tabular}%
  \label{tab:n2tjfcistData}%
\end{table}%

\section{Appendix B: Details for N$_2$ Optimizations}

The details for the optimizations behind the final energies in the main paper are presented below.
In every case of N$_2$, all flavors of pure gradient descent based optimization were run for 2000 iterations at 10,000 samples per iteration except for the random step size method, which was run for 10,000 iterations at 2000 samples per iteration.
The total sampling cost was then 20 million samples, half of the standard linear method's cost of 40 million over 40 steps.
Given the descent methods' tendency to plateau after a few hundred iterations and then lower the energy by only a few m$E_h$ afterward, we expect that running them longer to fully match or exceed the linear method's sampling effort would not yield a much better result in most cases.
We found that it was often advantageous to allow for different types of parameters to be given different initial step sizes.
Tables \ref{tab:n2tjfcieqStep} through \ref{tab:n2allstStep} list the step sizes for different descent optimizations in all cases of N$_2$.
The values can be cross-referenced with the energy results in earlier tables to see which choices were most effective.
Some amount of experimentation was necessary to build up intuition for what choices are effective, but we generally expect more nonlinear parameters such as those in the $F$-matrix and the orbitals to require smaller step sizes.
We also found that RMSprop benefited from using larger step sizes compared to other descent algorithms.
The energy may be significantly raised on early iterations, but tends to be quickly lowered and eventually brought to an improved result once enough gradient history has built up over more steps. 

The details of the different hybrid method optimizations are slightly more involved and are discussed separately here.
In all cases, the blocked linear method steps of the hybrid optimization used 5 blocks with 5 directions from sections of RMSprop to provide coupling to variables outside a block and retained 30 directions from each block to construct the final space for determining the parameter update.
These were also the settings given to the blocked linear method optimizations of N$_2$ that appear in the main text.
All hybrid optimizations used the RMSprop method for their AD sections with the hyperparameters $d=100$, $\rho = .9$ and $\epsilon = 10^{-8}$.
The step sizes used in the AD sections varied over the course of the hybrid optimizations.
We typically chose larger step sizes for the AD portion of the first macro-iteration in order to obtain more energy and parameter improvement at a low sampling cost before any use of the BLM and these are tabulated separately as "Hybrid-Initial".
The smaller step sizes reported for the rest of hybrid optimization were used in the later macro-iterations to avoid rises in the energy that might occur before a sufficient gradient history was accumulated.
We also list the step sizes in the long RMSprop optimization used to achieve the descent finalized energies.
These were sometimes larger than those for the AD sections of the initial hybrid optimizations because the descent finalization was long enough for any early transient rises in the energy to recover.

We now specify how the hybrid method sampling costs reported in Tables \ref{tab:n2alleqData} and \ref{tab:n2allstData} of the main paper were divided between AD and the BLM. In all parameter equilibrium N$_2$, Hybrid 1 consisted of 500 AD steps costing 3 million samples interwoven with 12 BLM steps that cost 2.4 million samples.
Hybrid 2 consisted of the same sequence of steps, but had an increased sampling effort of 6 million samples on descent and 2.4 million on BLM.
Hybrid 3 had a greatly increased sampling cost and consisted of 1400 AD steps for 11.2 million samples interwoven with 19 BLM steps costing 38 million.
For all parameter stretched N$_2$, Hybrid 1 used the same sequence of steps and sampling cost breakdown as Hybrid 1 for the equilibrium case.
Hybrid 2 consisted of 600 AD steps that cost 7 million samples and 15 BLM steps that cost 3 million.
Hybrid 3 also had 600 AD steps, now using 12 million samples, and 15 BLM steps, now using 30 million samples.
For all descent finalizations in N$_2$, we used 1000 steps of RMSprop at an additional cost of 10 million samples and took an average over the last 500 steps to obtain our reported energies and error bars.

Finally, we give the breakdown of the hybrid sampling costs in Tables \ref{tab:n2tjfcieqData} and \ref{tab:n2tjfcistData} of Appendix A.
For the equilibrium case, Hybrid 1 had 500 AD steps costing 1.5 million samples and 12 BLM steps costing 2.4 million samples.
Hybrid 2 had the same combination of steps and BLM cost as Hybrid 1 while the AD steps used 3 million samples.
Hybrid 3 used the same sequence of steps, but increased the AD and BLM sampling costs to 5 million and 24 million, respectively.
In the stretched case, the hybrid optimization used 500 AD steps with 3 million samples and 12 BLM steps costing 24 million samples.

\begin{table}[H]
\footnotesize
    \caption{Step sizes for TJFCI equilibrium N$_2$ optimizations.}
    \centering
    \begin{tabular}{lccc}
         Method & 2-Body TJF & 1-Body TJF & CI 
   \\ \hline
  RMSprop 1 & 0.01  & 0.01  & 0.005 \\
   RMSprop 2 & 0.05  & 0.05  & 0.01 \\
            &       &       &        \\ \hline
    AMSGrad 1 & 0.05  & 0.05  & 0.01 \\
    AMSGrad 2 & 0.05  & 0.05  & 0.05 \\
    AMSGrad 3 & 0.01 & 0.01 & 0.01\\
    AMSGrad 4 & 0.001 & 0.001 & 0.001\\
            &       &       &        \\ \hline
    ADAM 1 & 0.01  & 0.01  & 0.01\\
    ADAM 2 & 0.005  & 0.005  & 0.005  \\
      ADAM 3 & 0.001  & 0.001  & 0.001  \\
            &       &       &         \\ \hline
    Random 1 & 0.01 & 0.01 & 0.01\\
    Random 2 & 0.0005 & 0.0005 & 0.0005 \\
    &       &       &         \\ \hline
    Hybrid-Initial 1 & 0.1  & 0.1  & 0.01\\
    Hybrid-Initial 2 & 0.1  & 0.1  & 0.01  \\
      Hybrid-Initial 3 & 0.1 & 0.1  & 0.01  \\
        &       &       &         \\ \hline
     Hybrid 1 & 0.001  & 0.001  & 0.001\\
    Hybrid 2 & 0.005  & 0.005  & 0.005  \\
      Hybrid 3 & 0.005  & 0.005  & 0.005  \\
    \end{tabular}
    \label{tab:n2tjfcieqStep}
\end{table}

\begin{table}[H]
\footnotesize
    \caption{Step sizes for TJFCI stretched N$_2$ optimizations.}
    \centering
    \begin{tabular}{lccc}
         Method & 2-Body TJF & 1-Body TJF & CI 
      \\ \hline
        RMSprop 1 & 0.01  & 0.01  & 0.005 \\
      RMSprop 2 & 0.05  & 0.05  & 0.01 \\
            &       &       &        \\ \hline
    AMSGrad 1 & 0.05  & 0.05  & 0.01 \\
    AMSGrad 2 & 0.05  & 0.05  & 0.05 \\
    AMSGrad 3 & 0.01 & 0.01 & 0.01\\
            &       &       &        \\ \hline
    ADAM 1 & 0.01  & 0.01  & 0.01\\
    ADAM 2 & 0.005  & 0.005  & 0.005  \\
            &       &       &         \\ \hline
    Random 1 & 0.01 & 0.01 & 0.01\\
    Random 2 & 0.001 & 0.001 & 0.001 \\
    &       &       &         \\ \hline
    Hybrid-Initial 1 & 0.1 & 0.1 & 0.1\\
    Hybrid 1 & 0.005 & 0.005 & 0.005\\
    \end{tabular}
    \label{tab:n2tjfcistStep}
\end{table}
\begin{table}[H]
\scriptsize
    \caption{Step sizes for all parameter equilibrium N$_2$ optimizations.}
    \centering
    \begin{tabular}{lccccc}
         Method & 2-Body TJF & 1-Body TJF & $F$-Matrix & CI & Orbitals
  \\   \hline
    RMSprop 1& 0.005  & 0.005  & 0.001  & 0.001  & 0.001 \\
    RMSprop 2& 0.05  & 0.05  & 0.01  & 0.01  & 0.01 \\
          &       &       &       &       &  \\ \hline
    AMSGrad 1 & 0.05  & 0.05  & 0.005 & 0.01  & 0.001 \\
    AMSGrad 2 & 0.005 & 0.005 & 0.001 & 0.001 & 0.001 \\
          &       &       &       &       &  \\ \hline
    ADAM 1 & 0.05  & 0.05  & 0.005 & 0.01  & 0.001 \\
    ADAM 2 & 0.005 & 0.005 & 0.001 & 0.001 & 0.001 \\
          &       &       &       &       &  \\ \hline
    Random 1 & 0.001 & 0.001 & 0.001 & 0.001 & 0.001 \\
    Random 2 & 0.001 & 0.001 & 0.0005 & 0.001 & 0.0005 \\
    &       &       &       &       &  \\ \hline
    Hybrid-Initial 1 & 0.1 & 0.1 & 0.0001 & 0.01 & 0.01 \\
    Hybrid-Initial 2 & 0.1 & 0.1 & 0.0005 & 0.01 & 0.01 \\
    Hybrid-Initial 3 & 0.01 & 0.01 & 0.001 & 0.01 & 0.001 \\
    &       &       &       &       &  \\ \hline
    Hybrid 1 & 0.0001 & 0.0001 & 0.0001 & 0.0001 & 0.0001 \\
    Hybrid 2 & 0.001 & 0.001 & 0.0005 & 0.0005 & 0.0005 \\
    Hybrid 3 & 0.001 & 0.001 & 0.001 & 0.001 & 0.001 \\
    &       &       &       &       &  \\ \hline
    DF-Hybrid 1 & 0.001 & 0.001 & 0.001 & 0.001 & 0.001 \\
    DF-Hybrid 2 & 0.001 & 0.001 & 0.001 & 0.001 & 0.001 \\
    DF-Hybrid 3 & 0.001 & 0.001 & 0.001 & 0.001 & 0.001 \\
    \end{tabular}
    \label{tab:n2alleqStep}
\end{table}

\begin{table}[H]
\scriptsize
    \caption{Step sizes for all parameter stretched N$_2$ optimizations.}
    \centering
    \begin{tabular}{lccccc}
         Method & 2-Body TJF & 1-Body TJF & $F$-Matrix & CI & Orbitals
  \\   \hline
      RMSprop 1 & 0.05  & 0.05  & 0.05  & 0.01  & 0.01 \\
    RMSprop 2 & 0.1   & 0.1   & 0.01  & 0.01  & 0.005 \\
            &       &       &       &       &  \\ \hline
    AMSGrad 1 & 0.05  & 0.05  & 0.05  & 0.01  & 0.01 \\
    AMSGrad 2 & 0.05  & 0.05  & 0.01  & 0.02  & 0.001 \\
    AMSGrad 3 & 0.005 & 0.005 & 0.001 & 0.002 & 0.001 \\
            &       &       &       &       &  \\ \hline
    ADAM 1 & 0.05  & 0.05  & 0.05  & 0.01  & 0.01 \\
    ADAM 2 & 0.05  & 0.05  & 0.01  & 0.02  & 0.001 \\
    ADAM 3 & 0.005 & 0.005 & 0.001 & 0.002 & 0.001 \\
            &       &       &       &       &  \\ \hline
    Random 1 & 0.001 & 0.001 & 0.001 & 0.001 & 0.001 \\
    Random 2 & 0.001 & 0.001 & 0.0005 & 0.001 & 0.0005 \\
     &       &       &       &       &  \\ \hline
     Hybrid-Initial 1 & 0.1 & 0.1 & 0.01 & 0.01 & 0.001 \\
    Hybrid-Initial 2 & 0.1 & 0.1 & 0.01 & 0.01 & 0.001 \\
    Hybrid-Initial 3 & 0.1 & 0.1 & 0.01 & 0.01 & 0.001 \\
    &       &       &       &       &  \\ \hline
    Hybrid 1 & 0.0001 & 0.0001 & 0.0001 & 0.0001 & 0.0001 \\
    Hybrid 2 & 0.001 & 0.001 & 0.0005 & 0.0005 & 0.0005 \\
    Hybrid 3 & 0.001 & 0.001 & 0.0005 & 0.0005 & 0.0005 \\
    &       &       &       &       &  \\ \hline
    DF-Hybrid 1 & 0.001 & 0.001 & 0.001 & 0.001 & 0.001 \\
    DF-Hybrid 2 & 0.001 & 0.001 & 0.001 & 0.001 & 0.001 \\
    DF-Hybrid 3 & 0.001 & 0.001 & 0.001 & 0.001 & 0.001 \\
    \end{tabular}
    \label{tab:n2allstStep}
\end{table}

\section{Appendix C: Molecular Geometries}

\begin{table}[H]
     \caption{Structure of equilibrium styrene. Coordinates in \r{A}.}
     \centering
     \begin{tabular}{lrrr}
       \multicolumn{1}{l}{} &  &  & \\ \hline
 C     &    1.39295   &  0.00000    &  0.00000 \\
C      &    2.16042   & -1.19258    &  0.01801 \\
C      &    2.09421 &    1.23178    & -0.01914 \\
C       &   3.56585  &  -1.15969    &  0.05286 \\
C        &  3.50142   &  1.27211    &  0.01795 \\
C        &  4.23686   &  0.07503    &  0.06081 \\
C         & 0.00000   &  0.00000    &  0.00000 \\
C   &      -0.79515   & -0.93087    &  0.54406 \\
H    &      1.71222   & -2.11161    & -0.00239 \\
H     &     1.59237   &  2.12471    & -0.04753 \\
H      &    4.09818   & -2.03273    &  0.07153 \\
H       &   3.99086   &  2.16987    &  0.01692 \\
H   &       5.25794   &  0.10043    &  0.09503 \\
H    &     -0.46324   &  0.77112    & -0.42775 \\
H     &    -0.43431   & -1.72147    &  1.02278 \\
H      &   -1.78240   & -0.84577    &  0.49296 \\
       \hline
     \end{tabular}
     \label{tab:styreneGeo}
   \end{table}

\begin{table}[H]
     \caption{Structure of FNNF transition state. Coordinates in \r{A}.}
     \centering
     \begin{tabular}{lrrr}
       \multicolumn{1}{l}{} &  &  & \\ \hline
       N &  0.49939 & -0.44656 & -0.59377\\
       N & 0.57066 & 0.41224 &  0.5639 \\
       F &  -0.39084 & 0.14563 & -1.36959 \\
       F & -0.39807 & -0.12032 &  1.39157\\
       \hline
     \end{tabular}
     \label{tab:fnnfGeo}
   \end{table}
   
 \section{Appendix D: NCJF Gaussian Basis}  
   
  The form for the three dimensional Gaussian basis functions of NCJFs in the main text can equivalently be written as $g_j(\mathbf{r}) = \exp (\mathbf{r}^T \mathbf{A} \mathbf{r} - 2 \mathbf{B}^T\mathbf{r} +C)$ where $\mathbf{A}$ is a symmetric matrix defined by 6 parameters, $\mathbf{B}$ is a three-component vector, and $C$ is a single dimensionless number.
  We used 16 basis functions for the all parameter cases of N$_2$, 16 in styrene, and 4 in FNNF.
  Their complete specifications are presented in Tables \ref{tab:gaussianA}-\ref{tab:fnnfGaussian}.
  The components $A_{xx},A_{xy},A_{xz},A_{yy},A_{yz},A_{zz}$ with units of inverse square bohr are the same for each basis function within a particular system and are therefore listed separately in Table \ref{tab:gaussianA}.
  Tables \ref{tab:n2alleqGaussian}-\ref{tab:fnnfGaussian} contain components $B_x, B_y, B_z$, with units of inverse bohr and $C$ for each system's basis functions.

    \begin{table}[H]
\footnotesize
    \caption{Components of the matrix $\mathbf{A}$ for our systems.}
    \centering
    \begin{tabular}{lcccccc}
         System & $A_{xx}$ & $A_{xy}$ & $A_{xz}$ & $A_{yy}$ & $A_{yz}$ & $A_{zz}$
        \\   \hline
        Equilibrium N$_2$ & -6.9282 & 0.0 & 0.0 &  -6.9282 & 0.0  & -6.9282  \\
        Stretched N$_2$ & -6.9282 & 0.0 & 0.0 &  -6.9282 & 0.0  & -6.9282  \\
       Styrene &-0.1 & 0.0 &0.0 &-0.1 &0.0 &-0.1 \\
       FNNF  &-0.1 & 0.0 &0.0 &-0.1 &0.0 &-0.1 \\
    \end{tabular}
    \label{tab:gaussianA}
\end{table}
 
   \begin{table}[H]
\footnotesize
    \caption{Gaussian components for all parameter equilibrium N$_2$.}
    \centering
    \begin{tabular}{ccccc}
         Basis Function & $B_x$ & $B_y$ & $B_z$ & $C$
      \\   \hline
        $g_0$ &  -0.8 &-0.8 &-0.8& -0.2771 \\
        $g_1$ & 0.8 &-0.8 &-0.8 & -0.2771\\
        $g_2$ &-0.8 &0.8 &-0.8 & -0.2771\\
        $g_3$ & 0.8 &0.8& -0.8 & -0.2771\\
        $g_4$ & -0.8 &-0.8& 0.8 & -0.2771\\
        $g_5$ &0.8 &-0.8& 0.8  & -0.2771\\
        $g_6$ &-0.8 &0.8& 0.8  &-0.2771 \\
        $g_7$ &0.8& 0.8& 0.8 & -0.2771\\
        $g_8$ &-4.4787& -0.8& -0.8  &-11.2500 \\
        $g_9$ &-2.8787& -0.8 &-0.8  & -4.5981\\
        $g_{10}$ &-4.4787 &0.8& -0.8  & -11.2500\\
        $g_{11}$ &-2.8787& 0.8& -0.8  & -4.5981\\
        $g_{12}$ &-4.4787 & -0.8 & 0.8  & -11.2500 \\
        $g_{13}$ &-2.8787 & -0.8 & 0.8  & -4.5981\\
        $g_{14}$ &-4.4787& 0.8& 0.8  & -11.2500\\
        $g_{15}$ &-2.8787& 0.8& 0.8  & -4.5981\\
    \end{tabular}
    \label{tab:n2alleqGaussian}
\end{table}

  \begin{table}[H]
\footnotesize
    \caption{Gaussian basis functions for all parameter stretched N$_2$.}
    \centering
    \begin{tabular}{ccccc}
         Basis Function & $B_x$ & $B_y$ & $B_z$ & $C$
 \\   \hline
        $g_0$ &  -0.8 &-0.8 &-0.8& -0.2771 \\
        $g_1$ & 0.8 &-0.8 &-0.8 & -0.2771 \\
        $g_2$ &0.8& 0.8& -0.8  & -0.2771 \\
        $g_3$ &0.8 &0.8& -0.8 & -0.2771\\
        $g_4$ &-0.8 &-0.8 &0.8 & -0.2771\\
        $g_5$ &0.8 &-0.8& 0.8  & -0.2771\\
        $g_6$ &-0.8 &0.8& 0.8  & -0.2771\\
        $g_7$ &0.8 &0.8 &0.8 & -0.2771\\
        $g_8$ &-5.8015 & -0.8& -0.8 & -22.7322 \\
        $g_9$ &-4.2015 &-0.8 & -0.8  & -11.8474 \\
        $g_{10}$ &-5.8015 &0.8 &-0.8  & -22.7322\\
        $g_{11}$ & -4.2015& 0.8& -0.8 & -11.8474\\
        $g_{12}$ &-5.8015 & -0.8 & 0.8  & -22.7322\\
        $g_{13}$ &-4.2015 & -0.8 & 0.8  & -11.8474 \\
        $g_{14}$ &-5.8015 & 0.8 & 0.8  & -22.7322\\
        $g_{15}$ &-4.2015 & 0.8 & 0.8  & -11.8474\\
    \end{tabular}
    \label{tab:n2allstGaussian}
\end{table}

\begin{table}[H]
\footnotesize
    \caption{Gaussian basis functions for equilibrium styrene.}
    \centering
    \begin{tabular}{ccccc}
         Basis Function & $B_x$ & $B_y$ & $B_z$ & $C$
       \\       \hline
        $g_0$ & -0.2632 & 0.0 & 0.0 & -0.6929 \\
        $g_1$ & -0.4083 & 0.2254 & -0.003403& -2.1748\\
        $g_2$ &-0.3957 & -0.2328 &  0.003617  & -2.1081\\
        $g_3$ &-0.6738 & 0.2191 & -0.009989  & -5.0220\\
        $g_4$ &-0.6617 & -0.2404 & -0.003392  & -4.9561\\
        $g_5$ &-0.8007 & -0.01418 & -0.01149  & -6.4137\\
        $g_6$ & 0.0 & 0.0 & 0.0  & 0.0\\
        $g_7$ &0.1503 & 0.1759 & -0.1028  & -0.6409\\
        $g_8$ &-0.3236 & 0.3990 & 0.0004516  & -2.6392\\
        $g_9$ &-0.3009 & -0.4015 & 0.008982  & -2.5184\\
        $g_{10}$ &-0.7744 & 0.3841 & -0.01352  & -7.4750\\
        $g_{11}$ & -0.7542 & -0.41005 & -0.003197  & -7.3691 \\
        $g_{12}$ & -0.9936 & -0.01898  & -0.01796  & -9.8794 \\
        $g_{13}$ & 0.08754 & -0.1457 & 0.08083  & -0.3543\\
        $g_{14}$ & 0.08207 &  0.3253 &  -0.19328  & -1.4992 \\
        $g_{15}$ & 0.3368 & 0.1598 & -0.09316  & -1.4767 \\
    \end{tabular}
    \label{tab:styreneGaussian}
\end{table}

 \begin{table}[H]
\footnotesize
    \caption{Gaussian basis functions for FNNF.}
    \centering
    \begin{tabular}{ccccc}
         Basis Function & $B_x$ & $B_y$ & $B_z$ & $C$
   \\         \hline
        $g_0$ &  -0.09437 &0.08439 &0.1122& -0.2862 \\
        $g_1$ & -0.1078 & -0.07790 & -0.1066 & -0.2905 \\
        $g_2$ &0.07386& -0.02752& 0.2588  & -0.7320\\
        $g_3$ &0.07522 &0.02274& -0.2630  &-0.7533 \\
    \end{tabular}
    \label{tab:fnnfGaussian}
\end{table}






%

\end{document}